%% file: DDLab_ErroneousData_arXiv.tex
\documentclass[onecolumn]{DDLabJournalPaper}

\input{Preamble/DDLabIdentity}
\citationauthors{A. Manoj \emph{et al.}}

\makeatletter
\patchcmd{\DDJ@titleblock}{\LARGE\bfseries}{\Large\bfseries}{}{\PackageWarning{local}{title-size patch failed}}
\patchcmd{\DDJ@titleblock}{{\normalsize\DDJ@renderauthors\par}}
  {{\large\DDJ@renderauthors\par}\vspace{0.15em}}{}{\PackageWarning{local}{author-size patch failed}}
\makeatother

\manuscripttype{Perspective}
\venue{Preprint}
\preprintid{}                 
\manuscriptversion{}

\title{No Value Is Better Than a Wrong Value: Finding and Handling
       Erroneous Data in Literature-Mined Datasets for AI}
\shorttitle{Tackling erroneous data in literature-mined datasets for AI}

\DDAuthor{\mbox{\textbf{Achyut Manoj}}}{a,1}
\DDAuthor{\mbox{\textbf{Advai Aiyer}}}{b,1}
\DDAuthor{\mbox{\textbf{Andrew Lee}}}{c,1}
\DDAuthor{\mbox{\textbf{Nayan Saini}}}{d,1}
\DDAuthor{\mbox{\textbf{Pragya Birla}}}{e,1}
\DDAuthor{\mbox{\textbf{Dibakar Datta}}}{f,*}
\citationyear{2026}

\DDAffiliation{a}{Tenafly High School, Tenafly, NJ 07670, USA}
\DDAffiliation{b}{The Academy for Mathematics, Science \& Engineering, Rockaway,
  NJ 07866, USA}
\DDAffiliation{c}{Regis High School, New York, NY 10028, USA}
\DDAffiliation{d}{Department of Electrical and Computer Engineering, University
  of Illinois Urbana-Champaign, Urbana, IL 61801, USA}
\DDAffiliation{e}{Department of Electrical Engineering, New Jersey Institute of
  Technology (NJIT), Newark, NJ 07103, USA}
\DDAffiliation{f}{Department of Mechanical Engineering, New Jersey Institute of
  Technology (NJIT), Newark, NJ 07103, USA}
\DDAffiliation{1}{Equal contribution; research intern, Dibakar Datta Laboratory (DDLab), NJIT, Newark, NJ 07103, USA}

\correspondence{\href{mailto:ddlab@njit.edu}{ddlab@njit.edu}}
\date{\today}

\paperabstract{%
Machine learning in materials science, chemistry, energy, and medicine
increasingly relies on datasets built by reading the published literature,
because for many properties no curated database exists. Reading thousands of
papers automatically is never error-free. Text recognition misreads digits,
units and notations are confused, values are attached to the wrong label,
material, or condition, language models occasionally invent values, and errors
in one paper are copied into later ones. This Perspective asks how such errors
can be found and kept away from AI models. Drawing on our work building an electro-optic
property database from the literature, we present a taxonomy of errors
organised by where they arise and the signature they leave. We then describe a
layered detection framework that combines schema and provenance checks,
physical admissibility (bounds, crystal symmetry, and governing equations),
cross-source statistical consensus, model-based auditing, and review ordered by
suspicion. Four demonstrations test the framework. In the latest production run
of our pipeline, the verification gate stopped 402 of 816 candidate values and
recorded a reason for each. Even so, an audit against a curated reference showed that 30 of
the 112 values it accepted from three benchmark papers were defective, and
simple reference-free checks, chiefly symmetry rules applied to every crystal
class, caught 21 of them. On measured BaTiO$_3$ and KNbO$_3$ tensors,
constitutive identities caught eight of ten realistic single-value errors and
pointed to the corrupted value. In a controlled benchmark with known ground
truth, the combined layers caught 81\% of errors at 85\% precision, including
errors in materials reported only once. Cleaning brought the accuracy on held-out materials to within about 3\% of a
model trained on error-free data. Without cleaning, 89\% of the twenty largest
values in the data were errors. This matters for discovery, because an AI
search for the best material picks the largest values first. The real data also
show a second problem: once wrong numbers are under control, missing
measurement conditions become the main limit, because many of our
electro-optic records cannot be told apart by the conditions recorded with
them. We
close with directions for the field, including measured extractor error rates,
shared physics rule libraries, and databases that report their own quality.}

\highlights{%
\begin{itemize}[leftmargin=1.2em, itemsep=0.15em, topsep=0.1em]
  \item A taxonomy of errors in literature-mined data, built from errors caught
        in a real electro-optic database and matched to the detector that can
        see each one.
  \item A five-layer detection framework (schema, physics, consensus, model,
        review) that states what each layer can detect, what it cannot, and what
        it needs as input.
  \item In real production output, 27\% of accepted values from three benchmark
        papers were defective; reference-free checks caught 70\% of them.
  \item In a controlled benchmark, the combined layers found 81\% of errors at
        85\% precision and restored model accuracy, while uncleaned data put
        errors at the top of every ranking.
  \item Missing context, measured by the indistinguishable fraction $\Phi$, is
        the binding constraint once wrong numbers are under control.
\end{itemize}}

\plainsummary{%
To predict or design new materials, AI needs large amounts of reliable data,
and much of that data exists only in published papers. Reading thousands of
papers automatically is now possible, but some numbers are always read wrongly:
a decimal point is lost, a unit is confused, or a value is attached to the wrong
material. We show how to catch these mistakes using the laws of physics,
agreement between independent papers, and the AI model itself, and we show
that without these checks the ``best'' materials an AI suggests are often just
the biggest mistakes in the data.}

\newcommand{\rT}[1]{r^{T}_{#1}}
\newcommand{\Determined}{\textbf{Determined:}\hspace{0.35em}}
\newcommand{\NotDetermined}{\textbf{Not yet determined:}\hspace{0.35em}}
\makeatletter
\renewcommand{\PNote}[1]{\par\vspace{0.2em}\noindent{\small\color{DDInk}\itshape #1}\par}
\makeatother
\newcommand{\rS}[1]{r^{S}_{#1}}

\begin{document}
\maketitle

\section{Introduction}
\label{sec:intro}

Machine learning has changed how materials, molecules, and devices are
discovered, but the limit on what it can do is set more and more by the data it
learns from, not by the algorithm \citep{himanen2019, morgan2020}. Where large,
curated databases exist, progress has been dramatic. High-throughput
first-principles repositories such as the Materials Project \citep{jain2013},
OQMD \citep{saal2013}, and AFLOW \citep{curtarolo2012} made formation energies,
band structures, and elastic constants available at a scale no single laboratory
could produce. Most of the knowledge a researcher actually needs, though, is
still held in unstructured publications: the measured value of a functional
property, the conditions under which it was measured, the synthesis route that
produced the sample, and the device-level figure of merit
\citep{olivetti2020, kononova2021}. For a large and growing number of properties
no database exists at all. Anyone who wants to train a model on them must first
build the dataset, and in practice that means reading the literature
\citep{swain2016, court2018, huang2020, kim2017}.

Automated literature mining has become practical. Rule-based toolkits such as
ChemDataExtractor \citep{swain2016} have been used to build databases of
phase-transition temperatures \citep{court2018} and battery materials
\citep{huang2020}. More recently, large language models (LLMs) have extracted
structured records from scientific text with precision approaching that of
expert curators \citep{polak2024, dagdelen2024}. What these tools cannot do is
make every extracted value correct. A PDF is a page-layout format, not a data
format. Subscripts collapse into the main line, table cells come apart from
their headers, and scanned papers from before the 1990s reach the extractor as
images of text. Units, sign conventions, tensor notations, and measurement
conditions vary between papers. Some values are wrong in the original
publication. An LLM that reads the text well can still attach a number to the
wrong label, material, or condition, and it can produce a value that was never
in the text \citep{ji2023}.

\begin{KeyIdea}
Extraction at scale will never be perfect, so the realistic goal is not an
error-free database. The goal is a database whose errors are detected,
labelled, repaired where the repair can be verified, and otherwise kept from
reaching the model. That changes the question from ``how do we extract
perfectly?'', which cannot be solved at scale, to ``how do we find and
neutralise the values that are wrong?'', which can be answered with physics,
statistics, and the model itself.
\end{KeyIdea}

The need for such a strategy is not specific to materials science. The
cheminformatics community learned years ago that errors in public databases
reach published models: incorrect chemical structures
\citep{young2008, fourches2010, fourches2016} and inconsistent bioactivity values
combined from different sources \citep{kramer2012, landrum2024} measurably
degrade structure--activity models. The thermophysical-property community built
dynamic data evaluation around the same problem \citep{frenkel2005, chirico2013}.

In medicine, a meta-analysis combines the numbers reported by many clinical
trials into one overall answer. The numbers are copied by hand from the trial
reports, copying mistakes are common, and they can be large enough to change
the overall answer \citep{gotzsche2007, mathes2017}. Recent audits of widely used structure
databases for metal--organic frameworks found high error rates in supposedly
computation-ready structures \citep{white2025, gibaldi2025}.

In machine learning
more broadly, label errors in benchmark test sets change which model is ranked
best \citep{northcutt2021pervasive}. Work on data is often valued less than work on
models, and small data problems left early in a project then grow into larger
failures later, which \citet{sambasivan2021} call ``data cascades''.

\begin{figure}[tb]
  \centering
  \begin{tikzpicture}[
      every node/.style={font=\footnotesize, align=center},
      src/.style={ddevidence, text width=27mm, minimum height=9mm},
      lay/.style={ddmethod, text width=40mm, minimum height=9mm},
      outc/.style={ddresult, text width=24mm, minimum height=9mm},
      node distance=2.2mm]
    \node[src] (s1) {Source document\\typos, cited values};
    \node[src, below=of s1] (s2) {Conversion\\OCR, layout, scans};
    \node[src, below=of s2] (s3) {Extraction\\label, unit, material};
    \node[src, below=of s3] (s4) {Context\\lost conditions};
    \node[src, below=of s4] (s5) {Harmonisation\\conventions, duplicates};
    \node[lay, right=12mm of s1] (l0) {L0\enspace Schema, units, provenance};
    \node[lay, below=of l0] (l1) {L1\enspace Physics: bounds, symmetry,\\governing equations};
    \node[lay, below=of l1] (l2) {L2\enspace Cross-source consensus};
    \node[lay, below=of l2] (l3) {L3\enspace Model-based audit};
    \node[lay, below=of l3] (l4) {L4\enspace Targeted human review};
    \node[outc, right=12mm of l0] (o1) {Accept\\with weight};
    \node[outc, below=of o1] (o2) {Repair\\verified rescale};
    \node[outc, below=of o2] (o3) {Hold\\typed reason};
    \node[outc, below=of o3] (o4) {Down-weight\\in training};
    \node[ddcontrol, text width=24mm, minimum height=9mm, below=of o4] (o5) {Reject\\logged, not deleted};
    \node[above=1.5mm of s1, font=\small\bfseries] {Where errors arise};
    \node[above=1.5mm of l0, font=\small\bfseries] {Layered detection};
    \node[above=1.5mm of o1, font=\small\bfseries] {What is done};
    \draw[ddflow] ($(s3.east)+(0.5mm,0)$) -- ($(l2.west)-(0.5mm,0)$);
    \draw[ddflow] ($(l2.east)+(0.5mm,0)$) -- ($(o3.west)-(0.5mm,0)$);
    \draw[ddreturn] (l4.south) -- ++(0,-4mm) -| ($(l0.west)+(-4mm,0)$) -- (l0.west);
    \node[font=\scriptsize\itshape, anchor=north] at ($(l4.south)+(0,-4.6mm)$) {review decisions fed back as rules};
  \end{tikzpicture}
  \caption{Overview of the approach. Errors enter a literature-mined dataset at
  five points (left; \cref{sec:sources}). They are caught by five detection
  layers of increasing cost (centre; \cref{sec:detect}), and each value leaves
  with one of five outcomes (right). No value is silently deleted: every hold,
  repair, and rejection is recorded with its reason, so the database can be
  audited and the decision reversed.}
  \label{fig:overview}
\end{figure}
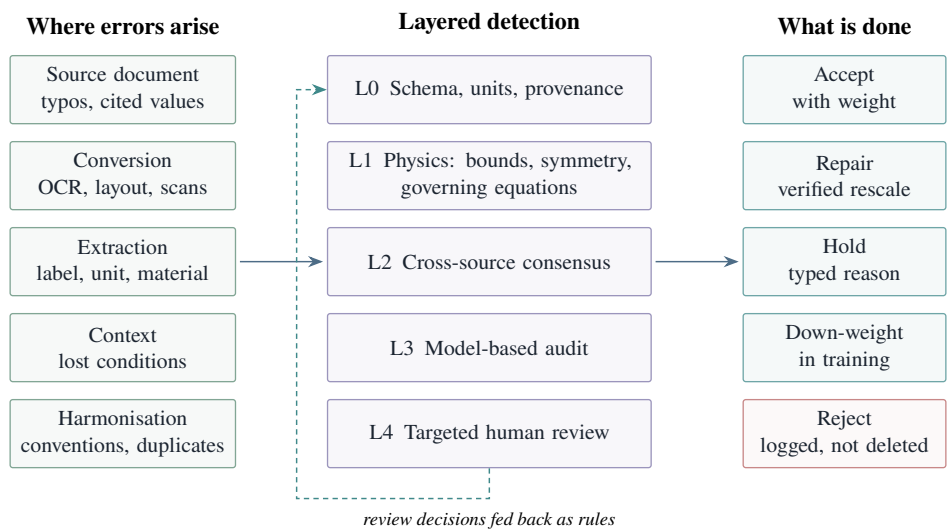

This Perspective asks how erroneous data should be handled when AI models are
trained on datasets built from the literature. We draw on our experience of
building an AI-ready database of electro-optic (Pockels) coefficients from the
literature \citep{datta2026framework, datta2026scikg}, where most of the
categories of error discussed below were met in practice. The argument is meant
to be general: the same failure modes appear wherever numbers are harvested from
publications, whether in battery research, medicine, or mechanics.
\Cref{fig:overview} summarises the approach.

The paper is a hybrid: part perspective, part quantitative demonstration. We
make four contributions.
\begin{itemize}[leftmargin=1.2em, itemsep=0.2em]
  \item A taxonomy of errors in literature-mined data, organised by where in the
        pipeline each error arises and by the signature it leaves in the data,
        illustrated with errors we caught in the electro-optic campaign
        (\cref{sec:sources}).
  \item A layered detection framework combining schema and provenance checks,
        physical admissibility and governing equations, cross-source statistical
        consensus, model-based auditing, and targeted human review. For each
        layer we state what it can detect, what it cannot, and what it needs as
        input (\cref{sec:detect}).
  \item Four demonstrations, two of them on real production data. We tally what
        the verification gate of our electro-optic pipeline stopped in its
        latest run and audit what it accepted against a curated reference; we
        show that constitutive identities on measured BaTiO$_3$ and KNbO$_3$
        tensors expose realistic single-value errors by many standard
        deviations; and in a controlled benchmark with known ground truth we
        measure the detection rate of each layer by error type and by source
        redundancy (\cref{sec:demos}).
  \item A quantitative account of the consequences for AI modelling: accuracy,
        tolerance to errors, and above all inverse design (searching for the
        material, or the conditions, that maximise a predicted property), which
        actively seeks out erroneous values (\cref{sec:impact}). The paper closes with
        directions for the field (\cref{sec:future}).
\end{itemize}

\noindent Because the paper draws on several fields, each symbol and
field-specific quantity is explained where it first appears, and all of them are
collected in \cref{app:notation}.

\section{AI-Ready Data Collection from the Literature}
\label{sec:collection}

\subsection{Why the literature, and why now}

Three conditions make literature mining necessary rather than merely convenient.
First, many functional properties are expensive or impossible to compute at
scale. Electro-optic, piezoelectric, and nonlinear-optical responses need
linear-response calculations with phonon and strain contributions. Ionic
conductivity, cycle life, and toxicity depend on processing and microstructure,
which no bulk calculation represents. The measured value in a paper is therefore
often the only value that exists. Second, a property value means something only
together with its context: the material and its form (bulk crystal, thin film,
nanostructure), the tensor component, the mechanical boundary condition, the
temperature, the frequency or wavelength, and the method. That context is
written in prose and table captions, not in a schema. Third, measured data on
rare or new materials are found only in papers. A composition that performs
unusually well has often been reported once, in a single paper, and appears in
no database.

The expected payoff is a database that is AI-ready, which means more than a
table of numbers. Each value is bound to its material, component, units, and
conditions. Each value carries its provenance (paper, page, table cell, raw
snippet). Each value has a confidence that later stages can use
\citep{wilkinson2016, gebru2021}. The rest of this paper is about the last
property: knowing which values can be trusted, and how far.

\subsection{Case study: electro-optic materials}
\label{sec:case}

Electro-optic materials serve here as a worked example; the approach does not
depend on them. An electro-optic material changes how it bends light when a
voltage is applied. That simple effect is how electrical signals are written
onto light: it drives the optical modulators that carry data through fibre
networks and data centres, and, at cryogenic temperatures, it can link
superconducting quantum processors to optical fibre
\citep{yariv1984optical, lauk2020perspectives, han2021microwave,
holzgrafe2020cavity, sahu2022quantum} (\cref{fig:fields}a). Our motivating
application is therefore the discovery of materials with a large linear
electro-optic (Pockels) response, which allows lower drive voltages and less
energy per bit. In the Pockels effect an applied electric field $E_k$ changes the
refractive index of a crystal in proportion to the field. The change is
described by the Pockels (electro-optic) tensor $r_{ijk}$, defined by
$\Delta(1/n^2)_{ij} = r_{ijk}E_k$, where $n$ is the refractive index, the
indices $i,j,k\in\{1,2,3\}$ label the crystal axes, and $r_{ijk}$ is usually
given in pm/V. It is a rank-three polar tensor whose allowed non-zero components
are fixed by crystal symmetry \citep{nye1985}.

Its value depends on whether the crystal is mechanically clamped or free, on
wavelength and temperature, and, in thin films, on strain and domain structure.
No curated database of measured Pockels coefficients exists. The values are
scattered across five decades of optics, ferroelectrics, and thin-film
literature, written in at least three notations: the full-tensor form $r_{ijk}$;
the compressed Voigt form $r_{ij}$, in which the symmetric index pair is replaced
by one index from 1 to 6 ($11\to1$, $22\to2$, $33\to3$, $23\to4$, $13\to5$,
$12\to6$); and the alternative symbol $\gamma_{ij}$ for the same quantity. A
superscript $S$ (constant strain, or clamped) or $T$ (constant stress, or
unclamped) states the mechanical boundary condition, and it is often dropped.
These and all other symbols are collected in \cref{app:notation}.

We built a pipeline to recover these values
\citep{datta2026framework, datta2026scikg} (\cref{fig:pipeline}). After an
acquisition step (Stage~0) finds the papers, four stages process them: (i) layout-aware extraction that
recovers tables and text together with the subject material and the measurement
conditions, using an LLM where contextual reasoning cannot be avoided; (ii) a
three-tier verification gate (accepted, review, rejected) that encodes symmetry
and provenance rules; (iii) cross-paper consensus with a robust rule based on the
median and the median absolute deviation (MAD; \cref{sec:consensus}); and (iv)
noise-robust modelling, audited by leave-one-material-out generalisation, in
which each material in turn is left out of training and then predicted. The pipeline
follows one rule: no value is better than a wrong value. Over the campaign,
staged runs over 49 distinct publications produced a 448-record consensus
database across 22 inorganic materials, and a forward surrogate (a
machine-learning model that predicts the property from descriptors of the
material and the measurement conditions) trained on it
generalised to nine of thirteen held-out materials \citep{datta2026scikg}.

\input{Figures/fig_pipeline}

The latest production run of this pipeline \citep{datta2026scikg} is the data
set analysed in \cref{sec:gate,sec:audit}. Forty-nine PDFs were processed with a
general-purpose LLM (GPT-4o) at a cost of under ten US cents per paper
\citep{datta2026scikg}. Three
papers were withheld at detection as out of scope (hybrid or organic
perovskites, or a multi-material survey), and 43 papers yielded 816 candidate
values. Of these, 414 were accepted, 215 were held for review, and 187 were
rejected, each with a typed reason. Two references let us measure how good the
result is. The first is a hand-curated set of tensor values from benchmark
papers \citep{zgonik1994, zgonik1993, jazbinsek2002}. The second is a certified
gold standard of 303 values from 16 papers, read by hand and independent of any
extraction pipeline \citep{datta2026scikg}. Against the gold standard, the production configuration
reaches a precision of 0.51 and a recall of 0.48 at the level of individual
target values. A diagnostic pass on an earlier run showed that numeric
precision was much higher (0.74) and that nearly half of the false positives
were the right number stored in the wrong slot: the wrong tensor label,
boundary condition, or contribution tier. These internal benchmarks are the
first measured error rates for our extractor, and \cref{sec:demos} builds on
them.

\begin{EvidenceNote}[What the campaign taught us]
\Determined{} every stage of the pipeline introduced errors that a
recall-first system would have passed into the database. In an early code
audit, two of the seven records promoted to the ``canonical'' set were
defective: one was a Pockels coefficient labelled as a piezoelectric
coefficient, and the other carried a temperature parsed from an electric-field
unit. Two months later, with a much better extractor, about half of the values
the pipeline emits still do not match the certified gold standard in every
field. \NotDetermined{} whether a stricter extractor alone can remove such
errors. Our experience is that it reduces them but does not remove them. Two
changes we tested and rejected make the point: adding a page-image reading pass
raised recall against the gold standard (0.37 to 0.43) but lowered precision
(0.46 to 0.38), and a prompt change meant to improve provenance tagging
collapsed recall \citep{datta2026scikg}. The remainder must be caught after extraction, which is what
the next two sections are about.
\end{EvidenceNote}

\subsection{The same need in other fields}

The electro-optic case is typical, not special. \Cref{tab:fields} lists domains
in which the property of interest is mostly reported in the literature, a
comprehensive database is missing or incomplete, and the reporting conventions
invite the same classes of error. What these fields share is the reason this
paper exists. Literature mining lets them build datasets that would otherwise
not exist. They then inherit error processes that the literature and the
extractor introduce together, and an unguarded model cannot tell those errors
from real physics.

\Cref{fig:fields} gives one schematic for each field of \cref{tab:fields}. In
\textbf{electro-optics}, a voltage across a crystal shifts the phase of light, and
the size of the shift is set by the Pockels coefficient. In \textbf{energy
storage}, ions move between two electrodes while electrons flow through the
external circuit, and the measured capacity is limited by how much charge the
electrode material can hold. In \textbf{drug delivery}, nanoparticles carry a
drug to target cells, and the reported numbers describe how much drug they hold
and how fast they release it. In \textbf{mechanics}, a tensile test pulls a
sample until it yields and breaks. In \textbf{superconductivity}, the electrical
resistance falls to zero below a critical temperature. In \textbf{thermophysics},
the pressure, temperature, and heat of a phase change are tied together by
thermodynamics. In every panel the same pattern holds: a number in a paper only
means something together with its conditions, and a law of physics limits which
numbers are possible.

\input{Figures/fig_fields}

\begin{DDWideTable}[tb]
  \caption{Fields in which AI models depend on data mined from the literature.
  For each field the table gives typical target quantities, the context that
  must be extracted with the value for it to be usable, and field-specific error
  hazards. Entries are representative, not exhaustive. Symbols and
  abbreviations are defined in \cref{tab:glossary} (\cref{app:notation}).}
  \label{tab:fields}
  \centering
  \small
  \begin{tabular}{L{0.15\linewidth} L{0.21\linewidth} L{0.24\linewidth} L{0.29\linewidth}}
    \toprule
    \DDHeadRow
    \DDHead{Field} & \DDHead{Typical targets} & \DDHead{Context that must travel with the value} & \DDHead{Characteristic hazards} \\
    \midrule
    Electro-optics, piezo- and ferroelectrics &
      $r_{ij}$, $d_{ij}$, $e_{ij}$, $\varepsilon_{ij}$, $n$ &
      Tensor component, clamped or unclamped, wavelength, temperature, strain, film or bulk &
      Voigt versus full-tensor factors of two, $\gamma$ versus $r$ notation,
      symmetry-forbidden components, \SI{e-11}{C/N} versus pm/V
      \citep{nye1985, zgonik1994, datta2026scikg} \\
    Energy storage &
      Capacity, Coulombic efficiency, cycle life, ionic conductivity, voltage &
      Loading, C-rate, electrolyte, temperature, cell format, reference electrode &
      mAh\,g$^{-1}$ versus mAh\,cm$^{-2}$, active-mass versus electrode-mass
      normalisation, S\,cm$^{-1}$ versus mS\,cm$^{-1}$, values read from plots
      \citep{huang2020, severson2019, goodenough2013} \\
    Medicine and drug delivery &
      IC$_{50}$, $K_\mathrm{i}$, encapsulation efficiency, release half-life, dose &
      Assay, cell line, species, dose basis, formulation, time point &
      nM versus \textmu M, pIC$_{50}$ versus IC$_{50}$, SD versus SEM,
      mg\,kg$^{-1}$ versus mg\,m$^{-2}$, arm-level versus patient-level values
      \citep{kramer2012, landrum2024, gotzsche2007, mitchell2021, altman2005} \\
    Mechanics and structural alloys &
      Yield and ultimate strength, hardness, fatigue life, toughness &
      Processing, grain size, strain rate, temperature, test standard &
      MPa versus GPa versus ksi, engineering versus true stress, HV versus GPa
      hardness, compression versus tension \citep{borg2020, hall1951} \\
    Superconductors and quantum materials &
      $T_c$, gap, critical field, coherence time &
      Doping, pressure, onset versus midpoint versus zero-resistance criterion &
      Criterion ambiguity, pressure-dependent values reported without the
      pressure \citep{stanev2018} \\
    Chemistry and thermophysics &
      Vapour pressure, density, solubility, reaction yield &
      Temperature, pressure, purity, phase &
      Thermodynamic inconsistency, unit and basis confusion (mole versus mass)
      \citep{frenkel2005, chirico2013} \\
    \bottomrule
  \end{tabular}
\end{DDWideTable}

\section{Sources of Error in Literature-Mined Data}
\label{sec:sources}

An error in a mined value is any difference between the stored record and what
the source actually established. Errors should be classified by the signature
they leave, not only by their cause, because the signature decides which
detector can see them. We group errors by the pipeline stage at which they arise
(\cref{fig:overview}, left) and illustrate most of them with errors met in the
electro-optic campaign (\cref{tab:taxonomy}; \citealp{datta2026scikg}).
\Cref{tab:record} first shows what a single record has to carry, and so how many
places there are for an error to enter.

\subsection{A taxonomy by pipeline stage}

\begin{table}[tb]
  \caption{What one literature-mined record has to carry, shown for an
  unclamped electro-optic coefficient of tetragonal BaTiO$_3$ as reported by
  \citet{zgonik1994}. Every field is a place where an error can enter; the last
  column names the error class of \cref{tab:taxonomy} that corrupts it.}
  \label{tab:record}
  \centering
  \small
  \begin{tabular}{L{0.19\linewidth} L{0.33\linewidth} L{0.38\linewidth}}
    \toprule
    \DDHeadRow
    \DDHead{Field} & \DDHead{Example} & \DDHead{How it can go wrong} \\
    \midrule
    Value and unit & 105 pm/V & OCR digit, lost decimal point, unit or scale slip \\
    Property and label & Pockels coefficient, $r_{333}$ ($r_{33}$ in Voigt notation) & Label swap ($r$ stored as $d$), symmetry-forbidden component \\
    Boundary condition & Unclamped ($T$, constant stress) & Superscript dropped; clamped and unclamped mixed \\
    Material and phase & BaTiO$_3$, tetragonal, point group $4mm$ & Comparison material attributed to the subject; wrong phase \\
    Sample form & Bulk single crystal & Film and bulk values mixed \\
    Conditions & Temperature, wavelength, strain, as stated in the paper & Condition lost or misparsed (``5\,kV/cm'' read as 5\,K) \\
    Uncertainty & Reported standard uncertainty & Index or subscript stored as uncertainty \\
    Origin & Measured in this paper, not cited & Cited value counted as a new measurement \\
    Provenance & Paper, table, cell, and verbatim snippet & Missing; the value cannot be checked or traced \\
    \bottomrule
  \end{tabular}
\end{table}

\paragraph{Source-level errors}
Some values are already wrong in the published paper, before any software
reads it: typographical errors, values miscopied from earlier work, figures whose axes
disagree with the text, and, rarely, fabricated data \citep{carlisle2017}. A
related and much more common hazard is the cited value: a paper quotes a
well-known constant from earlier literature in order to compare with its own
result. If the cited value is extracted as a new measurement, it is
double-counted, and it is often attached to the citing paper's own material and
conditions.

\paragraph{Conversion errors}
Turning a PDF into text damages information in well-documented ways
\citep{blecher2023, lopez2009}. Characters are confused by optical character
recognition (1 with 7, 3 with 8, 0 with O, 5 with 6). Decimal points and minus
signs are lost. Subscripts and superscripts collapse into the main line, so
$r_{33}$ becomes \Code{r33} and $\varepsilon^{S}_{11}$ becomes \Code{eS11}.
Two-column layouts are read across columns, table rows break apart, and scans of
older papers may give no usable text at all. In our own early pipeline, a
text-normalisation step meant to repair spacing instead broke words and formulas
apart, turning the journal name ``PHYSICAL REVIEW'' into ``PHYS ICAL REVI EW''
and the formula BaTiO$_3$ into ``Ba Ti O'', and every later stage worked on the
damaged text \citep{datta2026scikg}.

\paragraph{Extraction and interpretation errors}
Even on clean text, a number has to be bound to the right label, unit,
material, and component. Rule-based extractors bind by proximity, so they attach
numbers to whatever label is nearby. We found a tensor subscript (1133) stored
as the uncertainty of the preceding value, and phonon mode indices (``modes
4--6'') stored as a Pockels coefficient because the word ``Pockels'' was in the
same sentence \citep{datta2026scikg}. LLMs bind by meaning but make errors of their own. They swap
closely related quantities (an electro-optic coefficient $r$ labelled as a
piezoelectric coefficient $d$, which instead gives the strain produced by an
electric field), assign a comparison material's value to the paper's subject material, and
occasionally produce a plausible value that is not in the source at all
\citep{ji2023}.

\paragraph{Context errors}
The number can be right while the metadata attached to it is wrong. A
coefficient measured just above a phase transition, at cryogenic temperature,
or in a strained film may be ten or a hundred times its room-temperature bulk
value. If the condition is lost, a correct measurement becomes a wrong data
point. We found a temperature of ``5\,K'' that had been parsed from the phrase
``5\,kV/cm'' \citep{datta2026scikg}. In a later stage we found 51 strain-resolved records whose strain
labels had been dropped during harmonisation \citep{datta2026inverting}; the
values were intact, and what had been lost was the information that made them
usable. In the latest corpus, after the extraction was revised, the number of
records with an exact strain rose from 32 to 48 \citep{datta2026scikg}.

Recording conditions in more detail brings new hazards of its own. Once
the extractor was asked to keep conditions as the paper words them, and not
only as bare numbers, the step that turns those words back into numbers made
new mistakes. It turned ``below 50\,K'' into exactly 50\,K, ``partially
relaxed'' into a strain of exactly 0\%, and ``0.88\% biaxial compressive'' into
a positive (tensile) strain, and it filed an electrical state, ``partially
poled'', in the strain column \citep{datta2026scikg}.

The most important form of context loss cannot be seen by looking at one
record at a time. Many electro-optic records sit in groups that share the
material, tensor component, source paper, and every recorded condition, yet
carry different values (\cref{sec:consensus}). Each value may be a correct
measurement taken from the paper. What is missing is the condition that
explains why the values differ (for example the crystal phase, an applied
field, or the lattice constant of the film), because it was never recorded. A
model therefore sees identical inputs paired with different answers, and it
cannot learn from them.

\paragraph{Harmonisation errors}
Merging values from many papers requires common units and notations, and
conversion is itself a source of error. Unit conversions can fail silently: in
our first pipeline the unit label was changed while the number stayed the same.
Notation conventions differ by exact factors. In Voigt notation
(\cref{sec:case}) the shear piezoelectric coefficient $d_{15}$ equals twice the
full-tensor component $d_{131}$, because the Voigt shear strain counts the two
equal tensor components $d_{131}$ and $d_{113}$ together. A value moved between
the two conventions without the factor is therefore wrong by exactly two.

Symmetry folding brings a subtler hazard. Crystal symmetry makes some components
equal; in a tetragonal crystal, for example, the $x$ and $y$ axes are
equivalent, so $r_{23}=r_{13}$. Mapping such symmetry-equivalent labels to one
canonical label (here $r_{23}$ to $r_{13}$) is right for scoring, but in our corpus it also made 12 records count
the same measurement twice, because a paper listed both equivalent components
\citep{datta2026scikg}.

A record can also disagree with itself. In one elastic stiffness constant the stored text
field read 783\,GPa while the numeric field correctly held 7.83\,GPa, and for
five KNbO$_3$ stiffness constants the numeric field was a factor of 100 too
small after conversion from \SI{e11}{N\,m^{-2}} \citep{zgonik1993, datta2026scikg}. Duplicates arise when the same
paper is collected under two filenames or when one measurement is quoted in many
later papers.

\begin{DDWideTable}[tb]
  \caption{Taxonomy of errors in literature-mined data, with instances found
  during the electro-optic campaign \citep{datta2026scikg}, the signature each type leaves, and the
  detection layer that can see it (L0--L4; \cref{sec:detect}).
  ``Decade-quantised'' means the stored value differs from the true value by
  close to an integer power of ten.}
  \label{tab:taxonomy}
  \centering
  \small
  \begin{tabular}{L{0.13\linewidth} L{0.37\linewidth} L{0.26\linewidth} L{0.13\linewidth}}
    \toprule
    \DDHeadRow
    \DDHead{Error class} & \DDHead{Observed instance (electro-optic campaign)} &
    \DDHead{Signature in the data} & \DDHead{Detected by} \\
    \midrule
    Unit or scale & KNbO$_3$ piezoelectric $d$ reported in \SI{e-11}{C/N} stored
      as pm/V; stiffness a factor of 100 too small after conversion from
      \SI{e11}{N\,m^{-2}} \citep{zgonik1993} & Decade-quantised shift; often outside physical bounds
      & L0, L1, L2 (repairable) \\
    OCR digit and decimal & Formula and word fragments (``Ba Ti O''); in one table,
      ``rS131'' read as ``rS31'' and ``rS113'' as ``rS411'' & Arbitrary
      multiplicative shift, sometimes decade-quantised & L2, L3; L1 if coupled \\
    Label or component & Pockels $r$ stored as piezoelectric $d$; tensor subscript
      1133 stored as an uncertainty; mode indices stored as a coefficient & Value
      typical of a different quantity; violates a governing equation & L1, L3 \\
    Symmetry-forbidden & $r_{11}$, $r_{13}$, $r_{22}$, $r_{33}$ for zinc-blende InP
      \citep{krauser1986electro}; $r_{111}$ and $r_{222}$ for $mm2$ KNbO$_3$
      \citep{zgonik1993}; $d_{123}$ for $4mm$ BaTiO$_3$ \citep{zgonik1994}
      & Non-zero component forbidden by the point group & L1 (deterministic) \\
    Number look-alike & A CdTe ``$r_{41}=\SI{110}{pm/V}$'' that was the
      crystal direction $\langle110\rangle$ \citep{jeon2007electro} & Order-of-magnitude outlier against
      chemical neighbours & L1 bounds, L2, L3 \\
    Material misattribution & Relaxor 0.88\,Pb(Zn$_{1/3}$Nb$_{2/3}$)O$_3$--0.12\,PbTiO$_3$
      values assigned to PbTiO$_3$ \citep{lu2000linear}; every record of a batch stamped with a
      default material & Plausible magnitude, wrong identity; often in range &
      L2, L3; L0 provenance \\
    Context loss & ``5\,kV/cm'' parsed as 5\,K; ``below 50\,K'' stored as 50\,K;
      51 strain-resolved records with strain labels dropped; records
      indistinguishable by their recorded conditions & Correct number, wrong or
      missing condition; identical inputs with different values & L0
      completeness, L2 conflict ($\Phi$), L3 \\
    Notation convention & $\gamma_{ij}$ used for $r_{ij}$ (same quantity,
      different symbol); Voigt $d_{15}$ versus $d_{131}$ & Exact factor (1 or 2)
      or a false ``unknown label'' & L0 symbol map, L1 identities \\
    Cited value & GaAs $r_{41}$ quoted for comparison in a paper on another
      material & Duplicate of a canonical value, attached to the wrong paper &
      L0 provenance, L2 duplicates \\
    Hallucination & No confirmed case; three values held because the number was
      not verbatim in its source snippet & Plausible, in range, unsupported by
      any snippet & L0 snippet check, L3, L4 \\
    \bottomrule
  \end{tabular}
\end{DDWideTable}

\subsection{Random versus systematic errors}

For detection, the most useful distinction is whether an error is random or
systematic. A random error, such as a single OCR slip or a single mislabelled
cell, is independent of the errors in other papers. When several independent
sources report the same quantity, a random error lies away from the cluster of
correct values and is outvoted. A systematic error repeats: a unit convention
misread identically in every paper that uses it, or a wrong value propagated
through a chain of citations. Such errors can outnumber the correct values
locally and pull the consensus towards themselves. No purely statistical
detector can separate a systematic error that has become the majority from the
truth; that requires physics (\cref{sec:physics}) or provenance
(\cref{sec:schema}).

Two further regularities help. First, many errors are multiplicative: unit
prefixes, lost decimal points, and notation factors multiply the value by a
constant. In logarithmic space they become additive shifts, and many of those
shifts are quantised: close to an integer number of decades, or
$\log_{10}2\approx0.30$ for the Voigt factor of two. Second, many errors are structured by provenance. They cluster by
paper, by extraction route (text, table, or scanned image), and by publication
era, so a provenance tag is a strong prior on error probability.

\subsection{What to expect in other fields}

The same five stages produce errors specific to each field.

In \textbf{energy storage}, specific capacity is normalised by active mass in one paper and by
electrode mass or area in another, cycle life depends on the end-of-life
criterion, and many values are read off plots, which adds digitisation error
\citep{huang2020}.

In \textbf{medicine and drug delivery}, potencies such as the half-maximal
inhibitory concentration IC$_{50}$ are reported in nM, \textmu M, or as negative
logarithms (pIC$_{50}=-\log_{10}\mathrm{IC}_{50}$); dispersion is reported as the
standard deviation (SD) or the standard error of the mean (SEM), which differ by
a factor $\sqrt{n}$ for a sample of $n$ subjects \citep{altman2005}; and doses are
normalised by body mass or body surface area. Combining assays from different
sources adds variance of the size of the effects being modelled
\citep{kramer2012, landrum2024}, extraction errors in meta-analyses are frequent
\citep{gotzsche2007, mathes2017}, and automating the extraction speeds up
systematic reviews but does not by itself guarantee correctness
\citep{marshall2019, jonnalagadda2015}.

In \textbf{mechanics}, engineering and
true stress differ increasingly at large strain, hardness is reported on several
scales, and imperial units persist in older alloy data \citep{borg2020}.

In \textbf{structural and chemical databases}, missing hydrogens, disorder, and
residual solvent produce structures that are chemically impossible although they
parse as valid files \citep{white2025, chung2019}.

The details differ from field to field. The signatures do not, and the detectors of the next section are built
around signatures.

\section{How to Detect Erroneous Data}
\label{sec:detect}

No single detector sees every error class in \cref{tab:taxonomy}. We therefore
organise detection as a stack of layers (\cref{fig:overview}, centre), ordered by
cost and by what they need as input. The cheap, deterministic layers go first
and need no data beyond the record itself. The statistical and model-based
layers need many records. Human review, the most expensive layer, is kept for
the short list of values the other layers could not settle. Throughout, a record
is a value $v$ with its label, units, material, conditions, and provenance, and
we work in logarithmic space, $y=\log_{10}|v|$, wherever the quantity spans
decades, so that multiplicative errors become additive shifts. A shift of
$\Delta y=1$ is one decade, a factor of ten in $v$. All symbols are collected in
\cref{tab:symbols} (\cref{app:notation}). \Cref{tab:layers} summarises the five
layers before each is described in turn.

\begin{DDWideTable}[tb]
  \caption{The five detection layers at a glance. Layers are applied in order
  of cost; each one passes on a score or a typed reason, never a silent
  deletion.}
  \label{tab:layers}
  \centering
  \small
  \begin{tabular}{L{0.11\linewidth} L{0.17\linewidth} L{0.14\linewidth} L{0.14\linewidth} L{0.15\linewidth} L{0.13\linewidth}}
    \toprule
    \DDHeadRow
    \DDHead{Layer} & \DDHead{What it checks} & \DDHead{Rule or score} & \DDHead{Needs as input} &
    \DDHead{Catches} & \DDHead{Misses} \\
    \midrule
    L0 Schema and provenance & Label, unit dimension, required fields, verbatim snippet, own versus cited value &
      Controlled vocabulary; snippet match & One record and its source text &
      Malformed records, hallucinated values, cited values & Well-formed wrong numbers \\
    L1 Physics & Hard bounds, crystal symmetry, governing equations &
      $z_{\mathrm{phys}}$, \cref{eq:zphys}; \cref{tab:equations} & One record, or several values from one paper &
      Impossible values, forbidden components, unit slips, even in values reported once & Errors smaller than the measurement uncertainty \\
    L2 Consensus & Agreement between independent reports of the same quantity &
      Robust $z_i$, \cref{eq:mad}; $R_B$; $\Phi$, \cref{eq:phi} & Three or more reports &
      Random errors, decade shifts (repairable), mixed conditions & Copied errors that form a majority; values reported once or twice \\
    L3 Model audit & Agreement with what similar materials predict &
      Suspicion $s_i$, \cref{eq:suspicion} & A trained model and descriptors &
      Outliers among chemical neighbours, including singletons & Cannot tell an error from a real anomaly \\
    L4 Review & The short list the other layers cannot settle &
      Combined score $S_i$, \cref{eq:combined} & A person and the stored provenance &
      Whatever the evidence in the source decides & Limited by budget \\
    \bottomrule
  \end{tabular}
\end{DDWideTable}

\subsection{Layer 0: schema, units, and provenance}
\label{sec:schema}

The first layer checks that a record is well formed before asking whether it is
right. Is the label in the controlled vocabulary for its property family? Do the
units have the right dimension (a Pockels coefficient must be a length per
voltage)? Are the required context fields present, and is the value supported by
a verbatim snippet in the source? The snippet check is the cheapest defence
against hallucination: a value that cannot be found in the source text is held,
however plausible it looks. Symbol maps belong here too. Reading $\gamma_{ij}$
as $r_{ij}$, and mapping clamped and unclamped superscripts to a
boundary-condition field, recovered a complete five-component KTiOPO$_4$ tensor
that a literal reader would have discarded \citep{datta2026scikg}. Layer~0 also
records whether a value is the paper's own result or a cited one, and it removes
byte-identical duplicate documents. Finally, it checks that a record agrees with
itself: the value string, the parsed number, and the converted number must
describe the same quantity. In our corpus this simple test found stored values
that differed from their own source text by a factor of 100.

\subsection{Layer 1: physics}
\label{sec:physics}

Physics is the only layer that can reject a value without comparing it with
other data. That makes it the only protection for values reported once, and the
only protection against systematic errors that have become the majority. We
distinguish four strengths of physical constraint.

\paragraph{Hard admissibility bounds}
Many quantities have strict limits: a refractive index and a static relative
permittivity cannot be below one, the elastic stiffness matrix of a mechanically
stable crystal must be positive definite (all its eigenvalues positive)
\citep{mouhat2014}, the
Poisson ratio $\nu$ (lateral contraction per unit axial extension) of an isotropic
stable solid lies in $(-1, \tfrac12)$, and an efficiency cannot exceed its
thermodynamic limit. Below these are softer plausibility ranges set by the known
range of the property family. A value outside a hard bound is rejected; a value
outside a plausibility range is held.

\paragraph{Symmetry}
Neumann's principle requires every physical property tensor to be invariant
under the point group of the crystal, the set of rotations and reflections that
leave the crystal unchanged \citep{nye1985}. The allowed non-zero components are
therefore known in advance. Point groups are written here in the standard
Hermann--Mauguin notation: $\bar{4}3m$ is the cubic zinc-blende class (GaAs, InP,
CdTe), $4mm$ the tetragonal class of BaTiO$_3$ at room temperature, $mm2$ the
orthorhombic class of KNbO$_3$, and $3m$ the trigonal class of LiNbO$_3$. For the
zinc-blende class the only independent linear electro-optic component is
$r_{41}=r_{52}=r_{63}$ (Voigt notation, \cref{sec:case}). A
reported $r_{11}$ for InP is not a surprising measurement; it is an extraction
error, and it can be rejected deterministically. The same logic applies to
piezoelectric, elastic, and dielectric tensors and to any quantity constrained by
symmetry.

Two cautions come from our own corpus. First, the rule has to be
applied to every crystal class, not only the one that motivated it: a gate that
enforced the zinc-blende rule alone let forbidden labels for $mm2$, $4mm$, and
$3m$ crystals through (\cref{sec:audit}). Second, the allowed set depends on the
phase and on the axis setting, not on the material name alone. BaTiO$_3$ appears
in our corpus as tetragonal $4mm$ crystals, rhombohedral first-principles models,
and monoclinic films. For $3m$ crystals, the allowed shear-type components are
$r_{22}$ or $r_{11}$ depending on which axis the mirror plane is perpendicular
to. A symmetry violation under an unstated setting should be held for review,
not rejected.

\paragraph{Governing equations and constitutive identities}
The strongest physical check uses relations that connect several independently
reported quantities. Let $\vect{v}=(v_1,\dots,v_K)$ be a set of $K$ values
reported in one paper, or assembled for one material under matched conditions,
and let $g(\vect{v})=0$ be a physical identity that the true values satisfy
exactly. Because measured values carry uncertainty, $g$ evaluated on the reported
values is not exactly zero, and the question is whether it is further from zero
than the uncertainties allow. The normalised physics residual answers it:

\begin{KeyEquation}{Normalised physics residual}[Layer 1]
\begin{equation}\label{eq:zphys}
  z_{\mathrm{phys}} = \frac{g(\vect{v})}{\sigma_g}, \qquad
  \sigma_g^2 = \sum_k \left(\pd{g}{v_k}\right)^{2} \sigma_k^2 ,
\end{equation}
\end{KeyEquation}

\noindent
where $\sigma_k$ is the reported standard uncertainty of $v_k$ and $\sigma_g$
is the resulting standard uncertainty of $g$, obtained in practice by Monte
Carlo propagation: every input is redrawn many times from its uncertainty
distribution, and the spread of the resulting values of $g$ is taken as
$\sigma_g$ \citep{jcgm101}. A record set with $|z_{\mathrm{phys}}|>3$, a residual more
than three standard uncertainties from zero, contains at least one wrong value.
When one value appears in several identities, the pattern of violated identities
points to which value it is.

Electro-optic crystals, our case study, offer a set of such identities. A
crystal with a Pockels effect lacks a centre of symmetry, and every such crystal
is also piezoelectric: an electric field deforms it, and a deformation creates
an electric polarisation \citep{nye1985}. Because of this electromechanical
coupling, its electro-optic, dielectric, piezoelectric, and elastic constants
are not independent. A crystal can be measured either free to deform (unclamped, at constant stress,
superscript $T$) or held rigid (clamped, at constant strain, superscript $S$).
The two sets of electro-optic and dielectric tensors differ because a free
crystal also deforms piezoelectrically, and that deformation changes its optical
and dielectric response. The difference is fixed by the piezoelectric and
elasto-optic tensors,

\begin{equation}\label{eq:identities}
  \rT{ijk} = \rS{ijk} + p^{E}_{ijlm}\, d_{klm}, \qquad
  \varepsilon^{T}_{ij} = \varepsilon^{S}_{ij} + \frac{d_{ilm}\, e_{jlm}}{\varepsilon_0},
\end{equation}

\noindent
and the piezoelectric strain and stress tensors are related through the
compliance, $d_{kij} = e_{klm}\, s^{E}_{lmij}$ \citep{nye1985}. Here $d_{klm}$
is the piezoelectric strain coefficient (strain per unit electric field, in
pm/V), $e_{jlm}$ the piezoelectric stress coefficient (stress per unit field, in
C/m$^2$), $p^{E}_{ijlm}$ the elasto-optic tensor (change of $1/n^2$ per unit
strain), $\varepsilon_{ij}$ the relative permittivity (dielectric constant),
$\varepsilon_0$ the permittivity of vacuum, and $s^{E}_{lmij}$ the elastic
compliance, the inverse of the elastic stiffness $c^{E}_{ijlm}$. The superscript
$E$ means that these quantities are taken at constant electric field. Indices run
over the three crystal axes, and a repeated index is summed over. We test these
identities on real data in \cref{sec:identities}.

Similar relations exist in every field (\cref{tab:equations}). In chemistry,
charge and mass must balance. In batteries, the Nernst equation ties an
electrode potential to concentration \citep{bard2001}, and stored energy must
equal the integral of voltage over capacity. In thermophysics, the Gibbs--Duhem
equation links the chemical potentials of the components of a mixture, and the
Clapeyron equation ties the slope of a phase boundary to the latent heat
\citep{callen1985, frenkel2005}. In optics and electronics, the Kramers--Kronig
relations make the real and imaginary parts of a response function (for
example the refractive index and the absorption) determine each other
\citep{lucarini2005}. In transport, Onsager reciprocity requires the matrix of
coupled transport coefficients to be symmetric \citep{onsager1931}.
\Cref{tab:equations} collects these and other relations, with the error each
one exposes.

\begin{DDWideTable}[tbp]
  \caption{Governing equations and bounds that expose erroneous values, by
  field. ``Hard'' relations are exact, so a violation larger than the propagated
  uncertainty (\cref{eq:zphys}) means a wrong value; ``soft'' relations are
  empirical and justify review only. All symbols are defined in
  \cref{tab:glossary}.}
  \label{tab:equations}
  \centering
  \footnotesize
  \renewcommand{\arraystretch}{1.25}
  \begin{tabular}{L{0.12\linewidth} L{0.15\linewidth} L{0.31\linewidth} L{0.22\linewidth} L{0.05\linewidth}}
    \toprule
    \DDHeadRow
    \DDHead{Field} & \DDHead{Relation} & \DDHead{Equation} & \DDHead{A violation reveals} & \DDHead{Type} \\
    \midrule
    Electro-optics & Clamped and unclamped & $r^{T}_{ijk}=r^{S}_{ijk}+p^{E}_{ijlm}d_{klm}$ \citep{nye1985} & Boundary condition swapped; wrong $r$, $p$, or $d$ & hard \\
    Piezoelectricity & Strain and stress coefficients & $d_{kij}=e_{klm}\,s^{E}_{lmij}$ \citep{nye1985} & Unit slip in $d$ or $e$; Voigt factor of two & hard \\
    Dielectrics, optics & Index and permittivity & $\varepsilon=n^{2}$ at optical frequency; $\varepsilon\ge1$, $n\ge1$ & Mixed static and optical values; impossible value & hard \\
    Optics & Kramers--Kronig & $n(\omega)-1=\dfrac{2}{\pi}\,\mathcal{P}\!\int_0^\infty \dfrac{\omega'\kappa(\omega')}{\omega'^2-\omega^2}\,\mathrm{d}\omega'$ \citep{lucarini2005} & Index and absorption from different samples or wrong units & hard \\
    Crystals & Symmetry (Neumann) & Tensor invariant under the point group \citep{nye1985} & Forbidden component, wrong label & hard \\
    Elasticity & Mechanical stability & Stiffness matrix $c_{ij}$ positive definite; $-1<\nu<\tfrac12$ \citep{mouhat2014} & Mislabelled or rescaled constants & hard \\
    Elasticity & Voigt--Reuss bounds & $K_{R}\le K\le K_{V}$, $G_{R}\le G\le G_{V}$ \citep{hill1952} & Polycrystal modulus inconsistent with single-crystal data & hard \\
    Batteries & Faraday limit & $Q_{\mathrm{th}}=\dfrac{nF}{3.6\,M}$ mAh\,g$^{-1}$ ($M$ in g\,mol$^{-1}$) \citep{goodenough2013} & Wrong mass basis or material & hard \\
    Batteries & Energy and efficiency & $E=\int V\,\mathrm{d}Q$; Coulombic efficiency $\le100\%$ & Inconsistent energy, voltage, or capacity & hard \\
    Electrochemistry & Nernst & $E=E^{0}-\dfrac{RT}{zF}\ln Q_r$ \citep{bard2001} & Wrong reference electrode or temperature & hard \\
    Thermodynamics & Clapeyron & $\dfrac{\mathrm{d}P}{\mathrm{d}T}=\dfrac{L}{T\,\Delta v}$ \citep{callen1985} & Inconsistent vapour pressure and latent heat & hard \\
    Thermodynamics & Gibbs--Duhem & $\sum_i x_i\,\mathrm{d}\mu_i=0$ at fixed $T$, $P$ \citep{callen1985} & Inconsistent activity or mixture data & hard \\
    Transport & Onsager reciprocity & $L_{ij}=L_{ji}$ \citep{onsager1931} & Wrong coupled coefficient (e.g.\ Seebeck--Peltier) & hard \\
    Medicine & Summary statistics & $\mathrm{SEM}=\mathrm{SD}/\sqrt{n}$; $\mathrm{pIC}_{50}=-\log_{10}\mathrm{IC}_{50}$ \citep{altman2005} & SD and SEM swapped; nM and \textmu M confused & hard \\
    Chemistry & Conservation & Mass and charge balance; fractions in $[0,1]$ & Wrong stoichiometry or normalisation & hard \\
    Alloys & Vegard's law & $a(x)\approx(1-x)\,a_A+x\,a_B$ \citep{denton1991} & Wrong composition or phase & soft \\
    Mechanics & Hall--Petch & $\sigma_y=\sigma_0+k\,D^{-1/2}$ \citep{hall1951, cordero2016} & Grain size or strength in wrong units & soft \\
    Transport & Arrhenius & $\sigma T=A\exp\!\left(-E_a/k_BT\right)$ \citep{laidler1984} & Temperature in \textdegree C read as K; outlier point & soft \\
    Nonlinear optics & Miller's rule & $\chi^{(2)}\propto\chi^{(1)}\chi^{(1)}\chi^{(1)}$ \citep{miller1964} & Order-of-magnitude outlier & soft \\
    \bottomrule
  \end{tabular}
\end{DDWideTable}

\paragraph{Empirical laws as soft constraints}
Many relations hold approximately. Miller's rule states that a second-order
nonlinear optical coefficient is roughly proportional to the product of the
linear susceptibilities involved \citep{miller1964}. The Moss relation links a
high refractive index to a small band gap \citep{moss1985}. Vegard's law makes
the lattice parameter of an alloy vary linearly with composition
\citep{denton1991}, the Hall--Petch relation makes yield strength rise as the
inverse square root of grain size \citep{hall1951, cordero2016}, and Arrhenius
behaviour makes the logarithm of conductivity fall linearly with inverse
temperature \citep{laidler1984}. Their scatter is too large to reject a value, but large enough
deviations justify review. Such a law should be used with a threshold calibrated
on its known scatter, never as a hard filter, because the genuinely anomalous
material is exactly what a discovery campaign is looking for.

\begin{PhysicsInsight}
A physical constraint is the only detector whose power does not depend on
redundancy. Statistical consensus needs several reports of the same quantity. A
symmetry rule or a constitutive identity can reject a value that appears in only
one paper, and it cannot be outvoted by a systematic error that has propagated
through the literature. Physics is therefore the right first defence for the
long tail of rarely studied materials, which is where discovery happens.
\end{PhysicsInsight}

\subsection{Layer 2: cross-source statistical consensus}
\label{sec:consensus}

When the same quantity (same material, component, boundary condition, and
matched conditions) has been reported $n$ times, the correct values lie close
together around the true value, spread only by the usual differences between
laboratories, while random errors land far from this cluster
(\cref{fig:consensus}). The mean and the standard deviation are poor tools for
finding the outliers, because the outliers themselves pull them. For the $n$
log-values $y_1,\dots,y_n$ of a group we therefore use the median $\tilde{y}$
and the median absolute deviation (MAD). These barely change when a few values
are wildly wrong: more than half of the values in the group would have to be
wrong before the median could be pulled far from the cluster
\citep{iglewicz1993, leys2013, rousseeuw1993}:

\begin{KeyEquation}{Robust consensus score}[Layer 2]
\begin{equation}\label{eq:mad}
  z_i = \frac{0.6745\,(y_i - \tilde{y})}{\max(\mathrm{MAD},\, \delta_0)}, \qquad
  \mathrm{MAD} = \operatorname{median}_j |y_j - \tilde{y}| ,
\end{equation}
\end{KeyEquation}

\noindent
where $z_i$ is the robust score of record $i$ (\cref{fig:consensus}). The constant $0.6745$ makes
$z_i$ read like an ordinary $z$-score (a distance in standard deviations) when
the scatter is normally distributed. The floor $\delta_0$ encodes the
irreducible scatter between laboratories. Without it, a tight cluster of identical values (a common result of citation
copying) makes every legitimate value look like an outlier. Values with
$|z_i|>3.5$ are flagged. Four refinements make consensus much more useful in
practice.

\input{Figures/fig_consensus}

\paragraph{Decade-shift repair}
If a flagged value lies close to a whole number of decades from the median,
that is, $|(y_i-\tilde{y})-q|<\kappa$ for a non-zero integer $q$ and a small
tolerance $\kappa$ (0.12 decades in \cref{app:benchmark}), the error is almost
certainly a unit or decimal slip, and the value can be repaired as $y_i-q$ rather
than discarded. The same test with a shift of $\pm\log_{10}2$ in place of $q$
catches Voigt-factor errors. A repair is a
hypothesis, so it is recorded as a repair and never written over the raw value.

\paragraph{Consistency between sources}
Metrology has long faced the problem of combining measurements that do not agree
within their stated uncertainties. The Birge ratio \citep{birge1932} and the
scale factor used by the Particle Data Group \citep{navas2024},
$R_B=\sqrt{\chi^2/(n-1)}$ with $\chi^2=\sum_i (v_i-\bar{v})^2/\sigma_i^2$, measure
whether a group of $n$ reports is mutually consistent. Here $\sigma_i$ is the
reported uncertainty of report $i$ and $\bar{v}$ the mean weighted by
$1/\sigma_i^2$. For consistent reports $R_B\approx1$; $R_B\gg1$ indicates an outlier or an unrecognised difference in
conditions. Fundamental-constant evaluations use the same idea when weighting
discrepant inputs \citep{tiesinga2021}. Applied to literature-mined groups, a
large $R_B$ is a request to check whether the group actually mixes two
conditions (for example clamped and unclamped) before any value is called wrong.

\paragraph{Within-source conflicts and the indistinguishable fraction}
Consensus across papers has a counterpart within a single paper. When two
records share every recorded feature (material, component, boundary condition,
source, and conditions) but disagree in value, at least one of them is
mislabelled, or a distinguishing condition was never captured. Let $\Set{G}$ be
the set of groups of records with identical feature vectors, $|G|$ the number of
records in group $G$, and $N$ the total number of records. The share of records
in groups whose values disagree,

\begin{KeyEquation}{Indistinguishable fraction}[reference-free]
\begin{equation}\label{eq:phi}
  \Phi = \frac{1}{N}\sum_{G\in\Set{G}} |G|\;
  \mathbf{1}\!\left[\frac{\max_{i\in G}|v_i|}{\min_{i\in G}|v_i|} > 1+\eta\right],
\end{equation}
\end{KeyEquation}

\noindent
measures how much of a database a model cannot learn from, whatever its
algorithm. Here the indicator $\mathbf{1}[\cdot]$ equals 1 when the condition in
brackets holds and 0 otherwise, and $\eta$ is a tolerance of about 10\%. In
words, $\Phi$ is the fraction of records that have a ``look-alike'': another
record with the same material, component, source, and recorded conditions, but
a value that differs by more than $\eta$. A model receives the same input for
look-alike records and is asked for different outputs, so it cannot fit them
all. $\Phi$ needs no reference data, so it can be computed for any
literature-mined database. It should always be reported together with the list
of fields used to define a group, because adding a field can split a group and
lower $\Phi$. \Cref{tab:phi} gives $\Phi$ for the 232 accepted electro-optic
records of our latest production run under four choices of fields.

\begin{table}[tb]
  \caption{Indistinguishable fraction $\Phi$ (\cref{eq:phi}, $\eta=0.1$) of the
  232 accepted electro-optic records of the latest production run
  \citep{datta2026scikg}, for four choices of the fields that define a group.
  Every group also shares material, tensor component, and source paper. Fewer
  recorded fields means more look-alikes and a larger $\Phi$.}
  \label{tab:phi}
  \centering
  \small
  \begin{tabular}{L{0.62\linewidth} r}
    \toprule
    \DDHeadRow
    \DDHead{Fields used to define a group (besides material, component, source)} & \DDHead{$\Phi$} \\
    \midrule
    Every recorded condition and provenance field & 0.39 \\
    Temperature, strain, wavelength, and boundary condition & 0.47 \\
    Feature set of our modelling study \citep{datta2026inverting} & 0.54 \\
    Temperature and strain only & 0.72 \\
    \bottomrule
  \end{tabular}
\end{table}

\paragraph{Multivariate and distributional checks}
Some errors look normal in each property separately and abnormal only jointly,
for example a refractive index $n$ that does not match the reported optical
permittivity, although the two must satisfy $\varepsilon=n^2$ at optical
frequencies. Density- and isolation-based outlier detectors
\citep{breunig2000, liu2008, chandola2009} can find them in the joint space.
Digit-frequency tests such as Benford's law \citep{benford1938}, the observation
that the leading digits of many natural data sets follow a fixed, non-uniform
distribution, are sometimes
proposed for screening fabricated data; at the scale of a single material they
have little power, and we do not recommend them as a filter.

\begin{ModelingNote}
Consensus needs redundancy, and it fails in two predictable ways. With $n\le2$
reports it cannot tell which value is wrong. When a systematic error has been
copied into many reports, it votes for the wrong value. The number of
independent corroborating sources should therefore be stored with every
consensus record as a reliability field. A value supported by one paper and a
value supported by twenty should not enter training on equal terms.
\end{ModelingNote}

\subsection{Layer 3: model-based auditing}
\label{sec:modelaudit}

A model trained on the whole database learns the dominant, mostly correct,
relation between descriptors (numerical features of each material and
measurement, such as composition, structure, and conditions) and property.
Values the model cannot reproduce are then candidate errors. To avoid a model
simply memorising its own training labels, each record is scored with an
out-of-fold prediction $\hat{y}_{-i}$: the database is split into folds, and
record $i$ is predicted by a model trained on the other folds, so the model
never saw it:

\begin{KeyEquation}{Model-based suspicion}[Layer 3]
\begin{equation}\label{eq:suspicion}
  s_i = \frac{|y_i - \hat{y}_{-i}|}{1.4826\,\operatorname{median}_j |y_j - \hat{y}_{-j}|},
\end{equation}
\end{KeyEquation}

\noindent
and flagged when $s_i$ exceeds a threshold. The denominator, 1.4826 times the
median absolute residual, is a robust estimate of the standard deviation of the
residuals, so $s_i$ measures how many typical residuals record $i$ lies from its
prediction. The model borrows strength from
chemically similar materials, which is why this layer can flag a value reported
once, where consensus has nothing to compare it with. It is the regression form
of the idea behind confident learning and related label-error detectors
\citep{northcutt2021confident, brodley1999, frenay2014, song2023}. Ensembles, sets of models trained independently, add a second signal: where
members disagree strongly with each other, the label or the model is unreliable
\citep{lakshminarayanan2017}. Influence functions and data valuation
\citep{koh2017, ghorbani2019} rank records by how much each one changes the
fitted model, which is the right order in which to check them.

The limitation is equally clear. A model-based detector cannot tell an error
from a real anomaly, and a material that is truly exceptional is by definition
poorly predicted by its neighbours. Layer~3 must therefore never delete a value.
It ranks values for review and, if review is not possible, down-weights them in
training (\cref{sec:impact}).

\subsection{Layer 4: targeted human review}
\label{sec:review}

The cost of review must scale with the number of errors, not with the size of
the database. The layers above make this possible because each produces a score.
Dividing each score by its flagging threshold ($\tau_{\mathrm{phys}}$ for
physics, $\tau_c$ for consensus, $\tau_m$ for the model) and taking the largest
gives one combined suspicion score,
\begin{equation}\label{eq:combined}
  S_i=\max\bigl(|z_{\mathrm{phys},i}|/\tau_{\mathrm{phys}},\ |z_i|/\tau_c,\ s_i/\tau_m\bigr),
\end{equation}
so that $S_i>1$ means at least one layer flags record $i$. Reviewing records in
decreasing order of $S_i$ turns an impossible audit of
ten thousand records into a feasible audit of a few hundred. With provenance
attached (paper, page, table cell, raw snippet), each check takes seconds. The
reviewer's decisions are fed back as labels, so that the rules of Layers~0--1
improve and the same error is not reviewed twice. We quantify the efficiency of
this ranking in \cref{sec:benchmark}.

\subsection{What to do with a flagged value}

Detection is only half of the problem. A flagged value has five possible
outcomes (\cref{fig:overview}, right). It is \Term{accepted} with a confidence
weight if the flag is explained, for example by a genuine change of condition.
It is \Term{repaired} if the error is identified and the correction verified,
such as a decade shift or a Voigt factor. It is \Term{held} with a typed reason
if it cannot be decided. It is \Term{down-weighted} if it must be used before
review. It is \Term{rejected} if it violates a hard physical constraint. It is
never deleted without a trace. A database that stores only the values it
accepted cannot be audited, and it cannot tell a later acquisition step what is
missing. The complete procedure is stated in \cref{alg:protocol}
(\cref{app:protocol}).

\subsection{Domain-specific examples}

\paragraph{Electro-optic materials}
Layer~0 maps $\gamma_{ij}$ to $r_{ij}$ and the combined coefficient $r_c$ (a
single number that mixes several tensor components, as measured in some
interferometric set-ups) to an effective coefficient, and it separates intrinsic tensor components from
device-level effective values. Layer~1 applies point-group selection rules and
\cref{eq:identities}. Layer~2 groups by material, component, boundary condition,
and temperature range, because clamped and unclamped values can legitimately
differ by a factor of two or more \citep{zgonik1994}. Layer~3 audits against
composition descriptors. In an earlier run of our pipeline, a CdTe entry read
``$r_{41}=\SI{110}{pm/V}$'', far above any reported CdTe value; the same paper
reports $r_{41}=\SI{15.5}{pm/V}$ \citep{jeon2007electro}. Tracing the entry to
its source showed that ``110'' was the $\langle110\rangle$ direction along which
the samples were cleaved \citep{datta2026scikg}.

\paragraph{Batteries and energy storage}
Physical checks include the theoretical capacity from Faraday's law (a reported
specific capacity above the theoretical limit of the active material has an
error in normalisation or in material identity; \citealp{goodenough2013}), Coulombic efficiency (charge
delivered on discharge divided by charge stored on charge) at or below 100\% in
steady cycling, energy density equal to the integral of voltage over
capacity, and Arrhenius consistency of conductivities reported at several
temperatures. Consensus works well for widely reproduced systems such as
LiFePO$_4$ or graphite. For new electrolytes that are rarely reproduced,
Layers~1 and~3 carry the load.

\paragraph{Medicine and drug delivery}
Physical and logical checks include dose conversion between mg\,kg$^{-1}$ and
mg\,m$^{-2}$, encapsulation efficiency and drug loading between 0 and 100\%,
release fractions that do not decrease with time, and the internal consistency
of summary statistics. For example, the $\sqrt{n}$ relation between SD and SEM
identifies which one was reported \citep{altman2005}, and a mean that falls outside its reported
range reveals an error. Consensus between sources must respect the variability
between assays, which is known to be large for potency data
\citep{kramer2012, landrum2024}.

\paragraph{Mechanics}
Ultimate tensile strength cannot be below yield strength. Elongation and
reduction of area must be consistent. Hardness and strength are linked by
approximate empirical conversions that serve as soft constraints. Elastic
constants must give a positive-definite stiffness matrix. The modulus of a
polycrystal must lie between the Voigt and Reuss bounds, the averages of the
single-crystal constants computed under uniform strain and under uniform stress
\citep{hill1952} (the Voigt bound is unrelated to the Voigt index notation of
\cref{sec:case}).

\section{Demonstrations}
\label{sec:demos}

We test the framework of \cref{sec:detect} in four ways, moving from real data to
controlled data. First, we look at what the verification gate caught in our
latest production run. Second, we audit the values the gate accepted against a
curated reference, to see which errors survive the gate and which
reference-free checks would have caught them. Third, we show that governing
equations expose single-value errors in real tensor data. Fourth, we measure
detection rates by error type in a benchmark where every error is known.

\subsection{What the gate caught in a production run}
\label{sec:gate}

Of the 816 candidate values extracted in the latest production run
(\cref{sec:case}; \citealp{datta2026scikg}), 402 did not pass the verification gate. \Cref{tab:reasons}
groups the typed reasons attached to them. A value can carry more than one
reason, so the columns count every reason a value touches.

\begin{table}[tb]
  \caption{Typed reasons attached by the verification gate to held and rejected
  values in the latest electro-optic production run \citep{datta2026scikg} (816 candidate
  values from 43 papers: 414 accepted, 215 held, 187 rejected). A value can carry
  several reasons.}
  \label{tab:reasons}
  \centering
  \small
  \begin{tabular}{L{0.56\linewidth} r r r}
    \toprule
    \DDHeadRow
    \DDHead{Reason class} & \DDHead{Held} & \DDHead{Rejected} & \DDHead{Total} \\
    \midrule
    Corrupted, missing, or unrecognised tensor label        & 95 & 100 & 195 \\
    Out-of-scope property family (e.g.\ phonon frequencies) &  0 & 137 & 137 \\
    Cited or comparison value, not the paper's own result   & 54 &  32 &  86 \\
    Missing condition key (value from a series)             & 27 &   0 &  27 \\
    Family, label, or unit mismatch                         & 25 &   2 &  27 \\
    Implausible magnitude for the stated context            & 19 &   3 &  22 \\
    Symmetry-forbidden component (zinc-blende rule)         &  0 &  20 &  20 \\
    OCR corruption, or value not verbatim in the source     &  5 &  12 &  17 \\
    Not a measurement (equation number, imaginary mode)     &  0 &   4 &   4 \\
    \bottomrule
  \end{tabular}
\end{table}

Three observations matter for the rest of the paper. Label problems dominate:
nearly half of the stopped values carry a corrupted, incomplete, or
unrecognised label. Many of these are held under a strict rule that refuses any
label it cannot positively recognise, so the count mixes real errors with
unusual notation. Out-of-scope values are not errors; they simply fall outside
this database. Cited values are therefore the second largest class of genuine
hazard (86 values); a recall-first extractor would have double-counted each of
them under the citing paper's material and conditions. Finally, the one physics
rule that was implemented, zinc-blende symmetry, stopped 20 values on its own,
among them the components $r_{11}$, $r_{13}$, $r_{22}$, and $r_{33}$ extracted
for zinc-blende InP from a paper on GaInAsP/InP modulators
\citep{krauser1986electro}.
The next subsection shows what the same rule misses when it is not extended to
other crystal classes.

\subsection{Auditing accepted values against a curated reference}
\label{sec:audit}

A gate reports what it stopped, not what it let through. To measure what gets
through, we compared every accepted value from the three benchmark papers in the
corpus (BaTiO$_3$, KNbO$_3$, LiNbO$_3$; \citealp{zgonik1994, zgonik1993,
jazbinsek2002}) with our curated reference values for those papers. We then
applied five checks that need no reference data: Neumann's principle for every
crystal class (not only zinc-blende), label well-formedness, agreement between a
record's text and numeric fields, a plausibility range for diagonal stiffness
constants, and within-source conflict (\cref{eq:phi}). We added one constitutive
identity that the extracted values allow, the shear relation
$e_{131} = 2\, d_{131}\, c^{E}_{1313}$, which follows from $e=d\,c^{E}$ with
$c^{E}_{1313}$ the shear stiffness at constant field (the factor 2 again counts
the two equal shear components). \Cref{tab:audit} gives the result.

\begin{table}[tb]
  \caption{Accepted values from the three benchmark papers compared with the
  curated reference (112 records). ``Flagged'' counts records marked by at least
  one reference-free check: symmetry for all crystal classes, label
  well-formedness, text--number field agreement, stiffness plausibility,
  within-source conflict, or the shear identity.}
  \label{tab:audit}
  \centering
  \small
  \begin{tabular}{L{0.60\linewidth} r r}
    \toprule
    \DDHeadRow
    \DDHead{Outcome against the reference} & \DDHead{Records} & \DDHead{Flagged} \\
    \midrule
    Matches the reference (value within 2\%, component, boundary)   & 64 & 5 \\
    Value within 2--5\% of the reference                            &  4 & 1 \\
    Right number, wrong component label                             & 11 & 4 \\
    Right number, wrong boundary condition (clamped or unclamped)   &  2 & 2 \\
    Decade or unit shift                                            &  4 & 2 \\
    Component absent from the reference                             & 17 & 9 \\
    Value not found in the reference                                & 10 & 6 \\
    \midrule
    Total                                                           & 112 & 29 \\
    \bottomrule
  \end{tabular}
\end{table}

Only 64 of the 112 accepted values (57\%) match the reference in every field.
Seventeen are demonstrably wrong against it: eleven carry the right number under
the wrong component label, two carry the unclamped label on a clamped value, and
four are KNbO$_3$ piezoelectric coefficients reported in units of \SI{e-11}{C/N}
but stored as pm/V. Since \SI{1}{pm/V} equals \SI{e-12}{C/N}, the stored values
are ten times too small. A further 13 records are wrong by construction, whatever the
reference contains. Some carry labels that are forbidden or meaningless for
their crystal class ($r_{111}$ for $mm2$ KNbO$_3$, stiffness labels such as
$c_{1322}$ that orthorhombic symmetry forbids, $d_{123}$ for $4mm$ BaTiO$_3$,
and $d_{422}$, whose index 4 cannot occur in full-tensor notation). Others are five KNbO$_3$ stiffness constants whose numeric field
is a factor of 100 below their own source text. In total, 30 of the 112 accepted
records are demonstrably defective.

The reference-free checks flag 29 records. Twenty-four of these do not match the
reference, and the checks catch 21 of the 30 defective records. The remaining
five flags fall on correct values: four are the correct partners in a
within-source conflict, where the flag marks the group for review, and one is a
correct number whose text field was wrong by a factor of 100. The shear identity
is especially instructive. From the extracted $e_{131}=\SI{5.16}{C/m^2}$ and
$c^{E}_{1313}=\SI{25}{GPa}$ it implies $d_{131}=\SI{103}{pm/V}$. The extracted
value, \SI{10.73}{pm/V}, is smaller by a factor of 9.6. This is a
decade-quantised violation, so the check both detects the unit error and
proposes its repair. This audit covers three dense legacy tensor papers, one of
them scanned. It is a stress test, not a random sample of the corpus, and the
defect rate of a typical modern paper is likely lower.

\begin{KeyFinding}
In real production output, about a quarter of the values that passed a strict
gate were defective, and simple reference-free checks caught 70\% of them. The
nine defective values they missed are the ones physics cannot see without more
data: the right number under a label that is allowed for the crystal class, and
two piezoelectric unit errors for which the extracted elastic constants were
incomplete or mislabelled, so the identity could not be formed. Extending the
one implemented symmetry rule to every crystal class was the cheapest gain in
the whole study.
\end{KeyFinding}

\subsection{Constitutive identities expose extraction errors in real tensor data}
\label{sec:identities}

Complete, internally consistent sets of electromechanical and electro-optic
constants exist for a few benchmark ferroelectrics. We use the clamped and
unclamped dielectric and electro-optic tensors, the piezoelectric strain and
stress tensors, the elasto-optic tensor, and the elastic stiffness of tetragonal
BaTiO$_3$ (point group $4mm$) \citep{zgonik1994} and of orthorhombic KNbO$_3$
($mm2$) \citep{zgonik1993}, all as they appear in our curated reference set.
From these we evaluate sixteen identities of the forms in \cref{eq:identities}
and $d=e\,s^{E}$. Each identity has a normalised residual (\cref{eq:zphys}), with
$\sigma_g$ propagated from the reported uncertainties by Monte Carlo sampling:
every input is redrawn $2\times10^4$ times from a normal distribution with its
reported uncertainty, and $\sigma_g$ is the standard deviation of the resulting
values of $g$.

\begin{table}[tb]
  \caption{Constitutive identities evaluated on literature tensor data for
  BaTiO$_3$ \citep{zgonik1994} and KNbO$_3$ \citep{zgonik1993}. ``Reported'' is
  the value in the source; ``Implied'' is the value computed from the other
  reported tensors through \cref{eq:identities} or $d=e\,s^{E}$; $\sigma_g$ is
  the propagated uncertainty of the difference. Units: pm/V for $r$ and $d$
  (with $d_{15}$ in Voigt notation), dimensionless for $\varepsilon$.}
  \label{tab:identities}
  \centering
  \small
  \begin{tabular}{l l r r r r}
    \toprule
    \DDHeadRow
    \DDHead{Crystal} & \DDHead{Identity} & \DDHead{Reported} & \DDHead{Implied} &
    \DDHead{$\sigma_g$} & \DDHead{$z_{\mathrm{phys}}$} \\
    \midrule
    BaTiO$_3$ & $\rT{113}$                 &    8.0 &    8.0 &   3.2 & $+0.01$ \\
    BaTiO$_3$ & $\rT{333}$                 &  105.0 &  105.2 &  11.6 & $-0.02$ \\
    BaTiO$_3$ & $\rT{131}$                 & 1300   & 1294   & 185.4 & $+0.03$ \\
    BaTiO$_3$ & $\varepsilon^{T}_{11}$     & 4400   & 4378   & 485.9 & $+0.04$ \\
    BaTiO$_3$ & $\varepsilon^{T}_{33}$     &  129.0 &  129.4 &   8.9 & $-0.04$ \\
    BaTiO$_3$ & $d_{333}=(e\,s^{E})$       &   90.0 &   93.9 &  24.6 & $-0.16$ \\
    BaTiO$_3$ & $d_{311}=(e\,s^{E})$       &  $-33.4$ & $-33.7$ & 12.3 & $+0.03$ \\
    BaTiO$_3$ & $d_{15}=(e\,s^{E})$        &  564.0 &  560.7 &  57.0 & $+0.06$ \\
    KNbO$_3$  & $\rT{113}$                 &   33.5 &   33.7 &   3.4 & $-0.05$ \\
    KNbO$_3$  & $\rT{223}$                 &    6.0 &    6.3 &   1.6 & $-0.17$ \\
    KNbO$_3$  & $\rT{333}$                 &   63.3 &   63.5 &   3.2 & $-0.05$ \\
    KNbO$_3$  & $\rT{131}$                 &  120.0 &  120.5 &  23.4 & $-0.02$ \\
    KNbO$_3$  & $\rT{232}$                 &  450.0 &  448.9 &  48.5 & $+0.02$ \\
    KNbO$_3$  & $\varepsilon^{T}_{11}$     &  154.0 &  154.1 &   7.7 & $-0.01$ \\
    KNbO$_3$  & $\varepsilon^{T}_{22}$     &  985.0 &  986.1 &  31.0 & $-0.04$ \\
    KNbO$_3$  & $\varepsilon^{T}_{33}$     &   44.0 &   44.3 &   3.5 & $-0.08$ \\
    \bottomrule
  \end{tabular}
\end{table}

All sixteen identities hold, with $|z_{\mathrm{phys}}|\le0.17$
(\cref{tab:identities}). This is expected, because these data sets were derived
by self-consistent least-squares fitting \citep{zgonik1994, zgonik1993}, which
makes them a clean reference. The small $|z|$ values also show that treating the
reported uncertainties as independent overestimates $\sigma_g$. We keep that
conservative choice: it makes the test harder to trigger, so any violation it
reports is robust.

We then applied realistic single-value errors of the kinds catalogued in
\cref{tab:taxonomy}, one at a time, and re-evaluated all identities
(\cref{tab:injected}). Eight of the ten injected errors violate at least one
identity at $|z_{\mathrm{phys}}|>3$. The strongest violations reach
$|z|\approx15$--19 for decade and unit slips, and a clamped value stored in the
unclamped field gives $|z|=8.9$. The lost sign on $d_{31}$ and the Voigt factor
of two are caught through a different identity from the one we first expected,
because the same tensor element enters several relations. That redundancy is
useful: the set of violated identities points to the corrupted value. A wrong
$d_{333}$ in KNbO$_3$ violates exactly the four identities that contain
$d_{333}$; a wrong $\varepsilon^{S}_{22}$ violates only the identity for
$\varepsilon^{T}_{22}$.

\begin{table}[tb]
  \caption{Response of the identity check to single injected errors. Each row
  corrupts one value in the reference set. ``Max $|z|$'' is the largest
  normalised residual over all sixteen identities, and ``Fired'' counts the
  identities with $|z|>3$, which localises the corrupted value.}
  \label{tab:injected}
  \centering
  \small
  \begin{tabular}{L{0.42\linewidth} l l r r}
    \toprule
    \DDHeadRow
    \DDHead{Injected error} & \DDHead{Crystal} & \DDHead{Value} & \DDHead{Max $|z|$} & \DDHead{Fired} \\
    \midrule
    Decade shift (dropped decimal point)          & KNbO$_3$  & $d_{333}$              & 17.9 & 4 \\
    Unit: \SI{e-11}{C/N} read as pm/V             & KNbO$_3$  & $d_{131}$              & 15.2 & 2 \\
    Voigt $d_{15}$ stored as full-tensor $d_{131}$ & BaTiO$_3$ & $d_{131}$              & 10.0 & 2 \\
    Clamped value stored as unclamped             & KNbO$_3$  & $\rT{333}$             &  8.9 & 1 \\
    Sign lost                                     & BaTiO$_3$ & $d_{311}$              & 12.4 & 2 \\
    OCR digit, 9 read as 3 (90 stored as 30)      & BaTiO$_3$ & $d_{333}$              &  5.4 & 3 \\
    OCR digit, 7 read as 1 (780 stored as 180)    & KNbO$_3$  & $\varepsilon^{S}_{22}$ & 19.2 & 1 \\
    Stiffness unit slip (factor of ten)           & BaTiO$_3$ & $c^{E}_{33}$           & 17.0 & 2 \\
    Adjacent table cell read ($e_{311}$ value in $e_{333}$) & KNbO$_3$ & $e_{333}$     &  1.9 & 0 \\
    Small transcription slip (+10\%)              & BaTiO$_3$ & $\rT{333}$             &  0.9 & 0 \\
    \bottomrule
  \end{tabular}
\end{table}

\begin{KeyFinding}
Constitutive identities catch gross extraction errors in single-source data with
no comparison database: 8 of 10 realistic errors were caught, most at many
standard deviations, and the pattern of violated identities pointed to the
corrupted value. The two misses are errors that are small compared with the
uncertainty of the identity: a 10\% slip, and a swapped value whose contribution
to $\varepsilon_{33}$ is minor. This is the general limit of physics checks: they
are sensitive to the errors that do the most damage, and blind to errors smaller
than the combined measurement uncertainty.
\end{KeyFinding}

\subsection{A controlled benchmark with known ground truth}
\label{sec:benchmark}

Real literature data has no labels saying which values are wrong, so the recall
of a detector cannot be measured on it directly. We therefore built a synthetic
``literature'' in which every error is known, designed to reproduce the
statistical structure that matters for detection (\cref{app:benchmark}). Four
hundred materials are described by six descriptors, and their true property (in
$\log_{10}$) is a smooth nonlinear function of the descriptors plus a
material-specific term of 0.12 decades that no descriptor explains. The number of
papers per material is heavy-tailed, as in the real literature: just over half of
the materials are reported once, and about 6\% are reported more than ten times
(up to sixty). Each report carries between-laboratory and within-laboratory
scatter (together about 10\%). In 40\% of reports, two auxiliary quantities $u$ and $w$ are also reported,
related to the target by an exact identity, $10^{y}=u\,w$ (that is,
$y=\log_{10}u+\log_{10}w$). This is the analogue of reporting $r^{S}$, $p^{E}$,
and $d$ together with $r^{T}$ in \cref{eq:identities}. Errors from seven classes are then
injected so that a fraction $\rho$ of the records (the error rate) are wrong: unit or scale shifts, OCR digit and
decimal corruption, label swaps, misattribution to another material (often a
chemically similar one), context loss, hallucinated values drawn from the
distribution of the property, and one wrong value copied into several reports of
the same material, as through a chain of citations. We report means over ten
random corpora (about 1300 records each), for $\rho=15\%$ unless stated
otherwise.

\begin{figure}[tb]
  \centering
  \includegraphics[width=\linewidth]{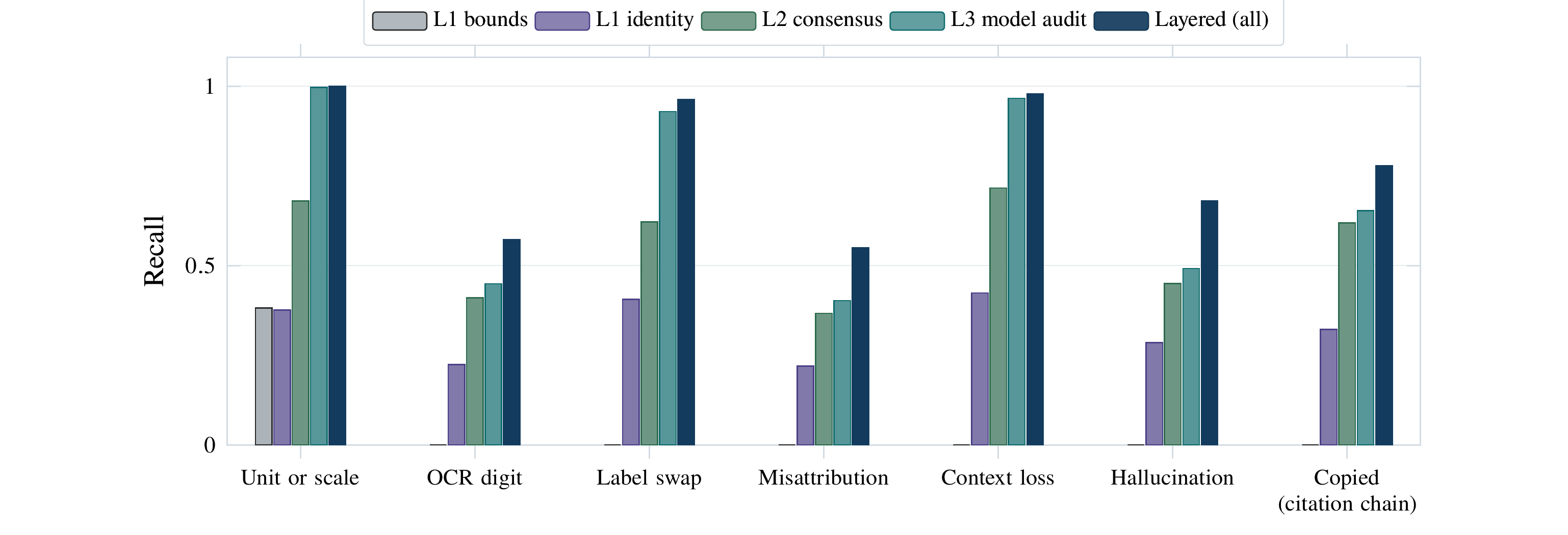}
  \caption{Recall by error class and detection layer in the controlled benchmark
  ($\rho=15\%$, mean of ten corpora). No single layer dominates: the physics
  checks are precise but cover few records, consensus needs redundancy, and the
  model audit has high recall at lower precision. Their union reaches 81\%
  recall at 85\% precision. The remaining errors are concentrated in the classes
  that produce plausible values.}
  \label{fig:recall}
\end{figure}

\paragraph{Each layer sees different errors}
\Cref{fig:recall} gives the recall of each detector by error class. The bounds
check is perfectly precise but catches only the 38\% of unit errors that leave
the plausible range. The identity check is also almost perfectly precise
(precision 0.998); it is limited by coverage, not by power, catching 82\% of the
errors in records that report the auxiliary quantities, which is 33\% of all
errors. Consensus catches 58\% of all errors with precision 0.98. The model-based
audit has the highest single-layer recall (0.72), but its precision is lower
(0.84), and it is the only layer that flags a noticeable fraction of clean
records (2.5\%). The layered combination catches 81\% of errors with precision
0.85. It catches every unit or scale error, 96--98\% of label swaps and context
errors, and 78\% of copied errors. The classes that remain hard are exactly the
ones that produce plausible values: misattribution to a similar material
(55\%), OCR digit corruption (57\%), and hallucination (68\%).

\begin{figure}[tb]
  \centering
  \includegraphics[width=\linewidth]{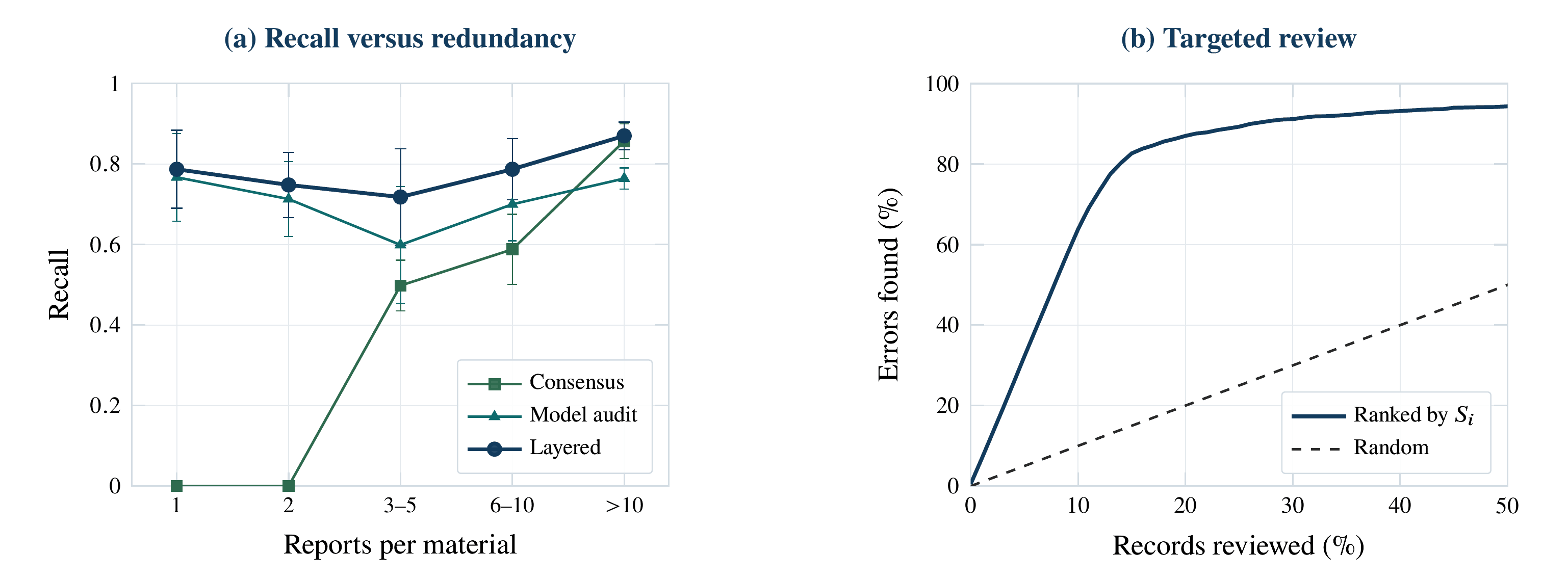}
  \caption{(a)~Recall as a function of how often the material is reported
  ($\rho=15\%$; bars are one standard deviation over ten corpora). Consensus
  cannot act below three reports, while the model audit, which borrows strength
  from similar materials, keeps the layered detector effective on the long tail.
  (b)~Fraction of all errors found when records are reviewed in order of the
  combined suspicion score, compared with random review.}
  \label{fig:redundancy}
\end{figure}

\paragraph{Redundancy decides what consensus can do}
\Cref{fig:redundancy}(a) splits recall by the number of reports of the material.
Consensus is blind for materials reported once or twice, and it reaches its best
performance (86\%) only for materials with more than ten reports. The
model-based audit is roughly flat across redundancy because it borrows strength
from chemically similar materials. As a result the layered detector keeps
75--79\% recall on the singletons and pairs that consensus cannot touch. This is
the quantitative form of the long-tail problem: the materials reported least
often are the most likely discoveries, and they are the materials for which
physics and models, not voting, must supply the checks.

\paragraph{Review effort scales with the number of errors}
\Cref{fig:redundancy}(b) orders all records by the combined suspicion score
$S_i$ (\cref{sec:review}) and plots the fraction of errors found against the
fraction of the database reviewed. With errors at 15\% of the records, reviewing
the top 10\% of the ranked list finds 64\% of all errors, and the top 15\% finds
83\%; random review would find 10\% and 15\%. For a database of ten thousand
records, the ranked list turns a review of every record into a review of about
fifteen hundred, most of which are actual errors.

\paragraph{Repair}
Of the values flagged by consensus as decade-quantised shifts, 92\% were
restored to within 0.1 decades of the truth by the repair rule of
\cref{sec:consensus}. The remaining 8\% were other errors that happened to fall
near an integer shift, which is why a repair is recorded as a hypothesis and not
written over the raw value.

\begin{Limitation}
The benchmark is synthetic. Its error mix, redundancy distribution, and noise
levels were chosen to resemble our electro-optic campaign, but they were not
estimated from it, and the absolute recall values depend on these choices. The
benchmark supports the relative conclusions: which layer sees which error class,
how consensus depends on redundancy, and how strongly ranking concentrates
errors. It does not support the specific percentages as predictions for any real
database. In particular, the descriptors explain the synthetic property up to a
0.12-decade term, which is favourable to the model-based audit; real materials
are less learnable. The real audit of \cref{sec:audit} found about 27\% of
accepted values defective in three dense, partly scanned tensor papers even after
the gate, which lies inside the 5--30\% range the benchmark explores, and it
found the same pattern the benchmark predicts: errors that produce plausible
values are the ones that survive. A larger hand-audited sample, across fields and
document types, is needed before the benchmark can be calibrated to real error
mixes (\cref{sec:future}).
\end{Limitation}

\section{Impact on AI Modelling}
\label{sec:impact}

Why does an error rate of a few percent matter to a model trained on thousands
of records? Because errors in extracted data are neither small nor symmetric,
and because the most valuable uses of a materials model, ranking and inverse
design, are exactly the uses that errors distort most.

\subsection{Why a few errors dominate training}

Under a squared-error loss, a value that is off by three decades contributes a
squared residual of 9 (in $\log_{10}$ units). With the scatter between and within
laboratories of our benchmark, a clean record contributes about
$0.043^2\approx1.9\times10^{-3}$. One unit error therefore adds as much to the
loss as roughly five thousand correct records, and its pull on the gradient
that drives training, which is proportional to the residual, equals that of about seventy. On a linear
scale the imbalance is larger still. Flexible models make this worse:
over-parameterised learners can fit arbitrary labels exactly \citep{zhang2017},
so with enough capacity an erroneous value is memorised rather than averaged
away, and it distorts predictions for its chemical neighbours. Robust losses
such as Huber and absolute error \citep{huber1964, friedman2001} limit the pull
of each record, but they cannot tell a wrong value from an unusual correct one.

\subsection{Accuracy: what cleaning recovers}

We trained a gradient-boosted surrogate \citep{friedman2001} on the benchmark
corpora under six strategies and evaluated it on held-out materials against
their true property values (\cref{fig:modelling}a; \cref{app:benchmark}). With no
errors, all strategies reach a mean absolute error (MAE) of 0.13--0.14 decades.
The naive model, trained with a squared loss on everything extracted, degrades
steadily: its MAE rises by 32\% at $\rho=15\%$ (0.175 against 0.132 decades) and
by 47\% at $\rho=30\%$. A robust loss alone recovers about half of the loss at
$\rho=15\%$ (0.153) and almost none at $\rho=30\%$ (0.190). Aggregating to one
median value per material, a common practice, gives almost no protection
(0.171), because most materials are reported once and their errors pass through
unchanged. Layered cleaning with repair of decade shifts recovers nearly all of
the gap: at $\rho=15\%$ it gives 0.136 decades, within about 3\% of the
error-free model, and it stays within 0.01 decades of the error-free model up to
$\rho=20\%$. Soft down-weighting by the suspicion score, which removes nothing,
performs almost as well (0.140) and is the safer choice when exceptional values
are expected.

Cleaning has a price, and it should be stated. With no errors present, the
model-based layer still flags 4.5\% of the (clean) records, and removing them
raises the MAE slightly, from 0.132 to 0.139 decades. Precision is lowest when
errors are rarest (0.55 at $\rho=5\%$, against 0.85 at $\rho=15\%$). A detector
tuned for a noisy corpus therefore needs to be re-tuned as the extractor
improves, which is one more reason to route model-based flags to review or to
down-weighting rather than to deletion.

The same effects appear in the real electro-optic corpus, measured with the
modelling code of our pipeline \citep{datta2026scikg}. When the consensus layer
was skipped, every record entered training with nearly the same weight, and the
surrogate's skill fell from 0.456 to 0.373, where skill is one minus the ratio
of the model's cross-validated MAE to that of a baseline that always predicts
the weighted mean. The weighted coefficient of determination $R^2$ fell from 0.626 to 0.569 on otherwise identical data.
When the model-based audit was run on the 232 accepted electro-optic records, it
flagged 20. About half of these were legitimate values, mostly electronic or
ionic contributions to a Pockels coefficient that were being modelled together
with total coefficients; the rest were worth checking against the source. This
is the behaviour the benchmark predicts for Layer~3: a useful triage list, with
about 50\% precision on its own, that must not be used to delete values.

\subsection{Evaluation: errors in test data hide the problem}

Errors damage the measurement of a model as well as its training. If the test
records are drawn from the same uncleaned database, their errors enter the
metric. In our benchmark at $\rho=15\%$, the naive and cleaned models have true
held-out MAEs of 0.175 and 0.136 decades. Measured against the uncleaned test
records, the apparent MAEs are 0.319 and 0.269: the metric nearly doubles, the
apparent improvement from cleaning shrinks from 22\% to 16\%, and the variation
between corpora grows three- to fivefold. This is the regression analogue of the
finding that label errors in test sets change which model is ranked best
\citep{northcutt2021pervasive}. It also means that a leave-one-material-out
audit, such as the one we use for the electro-optic surrogate
\citep{datta2026scikg}, is only as trustworthy as the held-out labels. In any
audit that claims generalisation, the test set should be cleaned first.

\subsection{Discovery: inverse design seeks out errors}

\begin{figure}[tb]
  \centering
  \includegraphics[width=\linewidth]{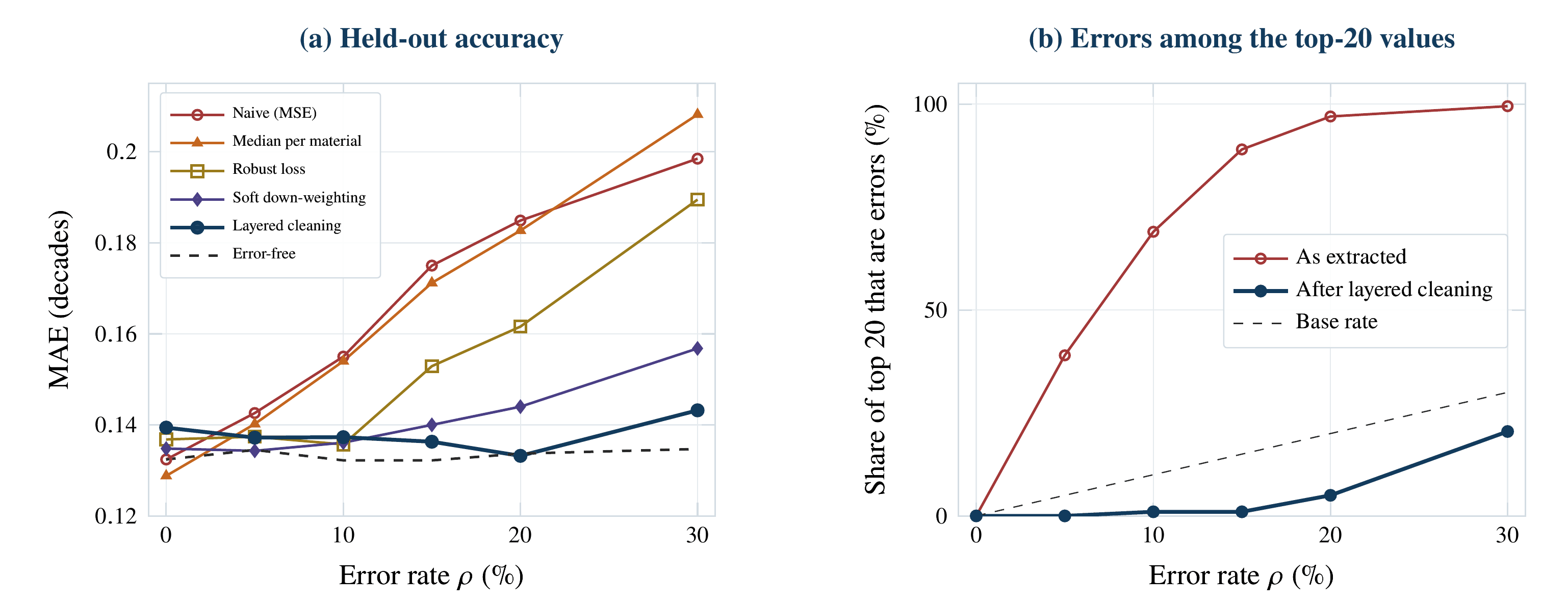}
  \caption{Consequences for modelling in the controlled benchmark (means over
  ten corpora). (a)~Mean absolute error on held-out materials, measured against
  the true values, for six training strategies. Layered cleaning with decade
  repair stays close to the error-free model up to $\rho=20\%$, whereas a robust
  loss alone or per-material median aggregation recovers only part of the loss.
  (b)~Share of errors among the twenty largest reported values, the records an
  inverse-design or screening step would select first. Errors are enriched far
  above their base rate (dashed line), because the largest errors are the
  largest values.}
  \label{fig:modelling}
\end{figure}

The most consequential effect concerns inverse design, and it follows from
simple selection. Screening and generative design look for the extremes of a
property, and the largest errors are, by construction, among the most extreme
values. \Cref{fig:modelling}(b) shows the result. When errors make up 15\% of
the records, 89\% of the twenty largest reported values are errors, and at
$\rho=5\%$ the share is already 39\%. After layered cleaning it falls to 1\% at
$\rho=15\%$. A discovery pipeline that ranks by the raw extracted value, or that
trains a surrogate and then maximises it, will therefore put errors at the top
of its candidate list, and it will do so with confidence, because the surrogate
has learned those values. This is the data-side version of the ``optimiser's
curse'' \citep{smith2006}: optimisation turns a random error into a systematic
bias. In our
inverse-design study for cryogenic Pockels materials, the surrogate was allowed
to use only the condition variables (such as strain and temperature) that the
data could support, and the verification step was
run on the database as well as on the candidates; that step exposed a
recoverable extraction defect that mattered more than any single ranked
candidate \citep{datta2026inverting}.

\begin{KeyTakeaway}
Erroneous data harms AI-driven discovery in three separate ways. It biases
training, since a single unit error can outweigh thousands of correct records.
It corrupts evaluation, since errors in test labels inflate the metric and hide
real improvements. Above all, it dominates the top of any ranking, since
optimisation selects extreme values and the largest errors are extreme values.
The third effect is the reason error handling must come before inverse design
and not be added afterwards.
\end{KeyTakeaway}

\subsection{The binding constraint is missing context}

The benchmark treats an error as a wrong number. In our real corpus the largest
obstacle to learning is different. Between 39\% and 72\% of the accepted
electro-optic records, depending on which recorded fields define a group, share
every recorded feature with another record that has a different value
(\cref{eq:phi}). No loss function, weighting scheme, or detector can recover what
distinguishes them, because it was never extracted. A related analysis found
that an extraction improvement that added records to one BiFeO$_3$ paper reduced
the usable information from that paper to zero, because the new values arrived
without the conditions that tell them apart \citep{datta2026inverting}. Record
counts cannot reveal this, and $\Phi$ can. We therefore recommend reporting
$\Phi$, with its feature set, together with precision against a gold standard,
as the two quality figures of any literature-mined database. When $\Phi$ is
large, the most valuable next step is to recover conditions, or to generate
condition-resolved data by calculation, not to mine more papers.

\subsection{Uncertainty and trust}

Two further effects are harder to quantify but matter for trust. First, a model
trained on data with errors reports uncertainties calibrated against labels that
include those errors, so its confidence is wrong in both directions: too wide
where errors add apparent scatter, and too narrow where a copied error has
produced false agreement. Ensembles \citep{lakshminarayanan2017} partly capture
the first effect and not the second. Second, the confidence weights,
corroboration counts, and typed reasons recorded by the detection layers are
themselves useful inputs to the model. They can serve as sample weights during
training, as features that let the model learn how reliable each source is, and
as the admissible-domain constraint of an inverse-design objective
\citep{datta2026scikg}. A database that carries its own quality information lets
the model decide how much to trust each record, instead of assuming that every
record is equally correct.

\section{Future Directions}
\label{sec:future}

The layered approach turns error handling from something done after the fact
into part of how the database is built. Several directions would make it
stronger and more widely used.

\paragraph{Measured error rates for real extractors}
The field lacks a basic number: how often does each extractor make each class of
error, on which kinds of source? We have taken a first step, with a certified
gold standard of 303 values from 16 papers and a reference audit of accepted
values (\cref{sec:case,sec:audit}; \citealp{datta2026scikg}). The measurements are sobering: precision
against the gold standard is about one half, and in three benchmark papers about
a quarter of the values that passed a strict gate were still defective. Much
larger hand-audited samples, stratified by publication era, document type
(table, text, scan), and field, would make it possible to calibrate detector
thresholds and to replace synthetic benchmarks like ours with real ones.
Publishing such audit sets alongside extraction benchmarks would do for
literature mining what curated test sets did for machine learning.

\paragraph{Encoding physics as reusable, machine-readable rules}
Symmetry selection rules, constitutive identities, and conservation laws are
known, but each project writes them again, often with different thresholds; our
own early pipeline had three inconsistent copies of its physical bounds. Shared
rule libraries, organised by property family and expressed so that both code and
paper can cite them, would give every project Layer~1 at no cost. The same
library could constrain LLM extraction directly: an extractor that knows
$r_{11}$ is forbidden for $\bar{4}3m$ crystals should never propose it.

\paragraph{Extraction that verifies itself}
LLM extractors can be asked to return, with each value, the verbatim span that
supports it, the reason for the label they chose, and the alternatives they
rejected. Checking the span against the source (Layer~0) and re-extracting in a
second independent pass make hallucination and label swaps detectable at the
point of extraction \citep{polak2024}. Disagreement between two extractors, or
between an extractor and a physics rule, is a cheap and informative review
signal.

\paragraph{Detecting systematic errors}
Consensus fails when an error has been copied through the literature. Tracing
citation graphs, so that a value quoted by ten later papers counts as one source
and not ten, and weighting groups of reports by the independence of their
measurements, would remove the main blind spot of statistical detection. The
same provenance graph also exposes clusters of papers that share an unusual
convention.

\paragraph{Databases that store uncertainty and provenance, not only values}
An AI-ready database should store, for each quantity, the individual reports,
their provenance, their quality flags, the consensus value, and a measure of how
consistent the reports are with each other \citep{birge1932, navas2024}. Held and
rejected values should be stored with typed reasons. Versioned releases, with
the record of each change, allow a model to be retrained on a corrected database
and let downstream users see what changed and why
\citep{wilkinson2016, gebru2021}.

\paragraph{Generating the missing conditions}
Where $\Phi$ is large, the missing information was usually never written down,
because the authors had no reason to tabulate it, and further mining cannot
recover it. First-principles calculations can generate the condition axis
(strain, temperature, field) at the density a model needs, with provenance
recorded by construction. How to weight dense computed data with a systematic
offset against sparse measurements is an open question; it probably requires
treating provenance as a level of a hierarchical model rather than as one more
feature.

\paragraph{Closing the loop with acquisition}
Detection results say where the data are weakest: which materials have a single
unconfirmed report, which groups disagree, which conditions are missing. Coupled
to uncertainty-guided acquisition \citep{lookman2019, settles2009}, the pipeline
could look for the paper, measurement, or calculation that would best confirm or
refute a suspect value. The same discipline applies at the corpus boundary: in a
first screen of the papers our acquisition stage retrieved, only about one in
ten reported an electro-optic coefficient at all \citep{datta2026scikg}.

\paragraph{Upstream: better reporting}
Many of the errors in \cref{tab:taxonomy} would disappear if property values
were published in machine-readable form, with units, tensor conventions, and
conditions stated explicitly. Journals already ask for data availability. Asking
for structured property tables in supplementary information, and checking them
against simple physical rules at submission, as has been done for thermophysical
data \citep{chirico2013}, would reduce errors where they begin.

\section{Conclusion}
\label{sec:conclusion}

Literature mining is becoming the main route to AI-ready data for properties
that are measured but not collected, and it brings errors that no extractor will
remove completely. We have argued that these errors should be treated as an
expected property of the data, not a failure to be hidden, and handled with a
layered approach. Schema and provenance checks catch malformed and unsupported
records. Physics, through bounds, symmetry, and governing equations, catches
errors even in values reported once. Statistical consensus catches random errors
wherever the literature is redundant. Model-based auditing extends coverage to
the long tail. Targeted human review, ordered by suspicion, settles what is left,
with effort that grows with the number of errors and not with the size of the
database.

Four demonstrations support this approach. In a production run of our
electro-optic pipeline, the gate stopped 402 of 816 candidate values with typed
reasons. Yet an audit against a curated reference showed that about a quarter of
the values it accepted from three benchmark papers were still defective, and
simple reference-free checks caught 70\% of them. On real BaTiO$_3$ and
KNbO$_3$ tensor data, constitutive identities caught eight of ten realistic
single-value extraction errors and pointed to the corrupted value. In a
controlled benchmark, the combined layers found about four fifths of injected
errors at 85\% precision, kept their power for materials reported only once, and
restored model accuracy to within a few percent of error-free training. The
benchmark also exposed the reason all this matters for discovery: without
cleaning, the extreme values that inverse design selects first are almost all
errors. The real corpus added a lesson the benchmark could not: once wrong
numbers are under control, missing context becomes the binding constraint, and it
has to be measured ($\Phi$) and repaired at the source.

The principle that governs our electro-optic pipeline also applies here: no value
is better than a wrong value. A database that knows which of its values to doubt
is worth more to AI-driven science than a larger database that treats every value
as correct.

\begin{Reproducibility}
The procedures behind \cref{sec:demos,sec:impact} are described in enough
detail to be reproduced. \Cref{app:benchmark} gives the benchmark generator, the
error model, the detector thresholds, and the model settings, and
\cref{app:audits} gives the matching rules and checks used in the real-data
audits. The production records of \cref{sec:gate,sec:audit} come from the
pipeline code cited in \cref{sec:case} \citep{datta2026framework, datta2026scikg}.
The one-off analysis scripts used for the audits and the benchmark are not part
of a released code archive.
\end{Reproducibility}

\begin{Declarations}
\AuthorContributions{%
  D.D. conceived the idea. All authors participated in coding, writing the
  manuscript, and reviewing.}
\Funding{%
  This work was supported by the National Science Foundation through the
  Designing Materials to Revolutionize and Engineer our Future (DMREF) program,
  award DMR-2522898. Any opinions, findings, and conclusions expressed are those
  of the authors and do not necessarily reflect the views of the sponsor.}
\DataAvailability{%
  The accepted, held, and rejected records of the latest production run,
  with their typed reasons, are the data of \cref{sec:gate,sec:audit}. The
  reference tensor values of \cref{sec:audit,sec:identities} are taken from the
  cited primary publications. The benchmark of \cref{sec:benchmark} is
  entirely synthetic and fully specified in \cref{app:benchmark}. The production records
  are available from the corresponding author on reasonable request.}
\CodeAvailability{The extraction, verification, consensus, and modelling
  pipeline used for the production run is described in
  \citet{datta2026framework} and \citet{datta2026scikg}; information on code
  access is available at \DDJCodeLink. The analysis scripts for the audits and
  the benchmark are not released; their procedures are given in
  \cref{app:benchmark,app:audits}.}
\CompetingInterests{The authors declare no competing interests.}
\AIDisclosure{%
  A general-purpose large language model (GPT-4o, OpenAI) performed
  candidate-value extraction inside the pipeline studied in
  \cref{sec:gate,sec:audit}. All
  quantitative results were checked by the
  authors, who are responsible for all content.}
\Acknowledgments{%
  The authors thank the DDLab members and the DMREF team for discussions.
  Computations used the NJIT high-performance computing facilities and ACCESS
  allocations DMR180013 and MAT250009.}


\end{Declarations}

\bibliographystyle{unsrtnat}
\bibliography{references}

\DDAppendices
\section{Notation and glossary}
\label{app:notation}

This appendix collects the symbols used in the paper, so that a reader from any
of the fields discussed can look up a quantity without searching the text.
\Cref{tab:symbols} lists the symbols of the detection framework and the
benchmark. \Cref{tab:glossary} lists the physical quantities and field-specific
terms, most of which appear in \cref{tab:fields,tab:taxonomy}.

\begin{table}[!htbp]
  \caption{Symbols of the detection framework (\cref{sec:detect}) and the
  benchmark (\cref{sec:benchmark,app:benchmark}). ``Decade'' is one unit of
  $\log_{10}$, a factor of ten.}
  \label{tab:symbols}
  \centering
  \small
  \begin{tabular}{L{0.17\linewidth} L{0.62\linewidth} L{0.13\linewidth}}
    \toprule
    \DDHeadRow
    \DDHead{Symbol} & \DDHead{Meaning} & \DDHead{Where} \\
    \midrule
    $v$, $v_i$ & A reported value (of record $i$), in the units of its property & \Cref{sec:detect} \\
    $y$, $y_i$ & Logarithm of a value, $y=\log_{10}|v|$, in decades & \Cref{sec:detect} \\
    $\Delta y$ & A shift in $y$; $\Delta y=1$ is a factor of ten in $v$ & \Cref{sec:detect} \\
    $\vect{v}$, $K$ & A set of $K$ related values $(v_1,\dots,v_K)$ entering one identity & \Cref{eq:zphys} \\
    $g(\vect{v})$ & A physical identity, equal to zero for the true values & \Cref{eq:zphys} \\
    $\sigma_k$, $\sigma_g$ & Standard uncertainty of value $v_k$, and of $g$ after propagation & \Cref{eq:zphys} \\
    $z_{\mathrm{phys}}$ & Normalised physics residual, $g/\sigma_g$ & \Cref{eq:zphys} \\
    $n$ & Number of reports in a consensus group & \Cref{sec:consensus} \\
    $\tilde{y}$ & Median of the $y_i$ in a group & \Cref{eq:mad} \\
    MAD & Median absolute deviation, $\operatorname{median}_j|y_j-\tilde{y}|$ & \Cref{eq:mad} \\
    $\delta_0$ & Lower limit on the MAD: the irreducible scatter between laboratories & \Cref{eq:mad} \\
    $z_i$ & Robust consensus score of record $i$ & \Cref{eq:mad} \\
    $q$, $\kappa$ & Integer number of decades in a decade-shift repair, and its tolerance & \Cref{sec:consensus} \\
    $R_B$, $\chi^2$ & Birge ratio and chi-squared statistic of a group of reports & \Cref{sec:consensus} \\
    $\bar{v}$, $\sigma_i$ & Uncertainty-weighted mean of a group; reported uncertainty of report $i$ & \Cref{sec:consensus} \\
    $\Phi$ & Indistinguishable fraction: share of records in groups with identical features but different values & \Cref{eq:phi} \\
    $\Set{G}$, $G$, $|G|$ & Set of groups with identical feature vectors; one group; its size & \Cref{eq:phi} \\
    $N$ & Total number of records & \Cref{eq:phi} \\
    $\eta$ & Tolerance within which two values count as equal (about 10\%) & \Cref{eq:phi} \\
    $\mathbf{1}[\cdot]$ & Indicator: 1 if the condition holds, otherwise 0 & \Cref{eq:phi} \\
    $\hat{y}_{-i}$ & Out-of-fold prediction of $y_i$ by a model that did not see record $i$ & \Cref{eq:suspicion} \\
    $s_i$ & Model-based suspicion score of record $i$ & \Cref{eq:suspicion} \\
    $S_i$ & Combined suspicion score, the largest threshold-normalised score & \Cref{eq:combined} \\
    $\tau_{\mathrm{phys}}$, $\tau_c$, $\tau_m$ & Flagging thresholds for the physics, consensus, and model scores & \Cref{eq:combined} \\
    $\rho$ & Error rate: fraction of records that are wrong & \Cref{sec:benchmark} \\
    $M$ & Number of materials in a benchmark corpus (400) & \Cref{app:benchmark} \\
    $\vect{x}$, $x_1,\dots,x_6$ & Descriptor vector of a material and its components & \Cref{app:benchmark} \\
    $y^{*}$, $\delta_m$ & True property of a material (in $\log_{10}$) and its unexplained material-specific part & \Cref{app:benchmark} \\
    $n_m$, $P$ & Number of reports of material $m$, and the heavy-tailed random number that sets it & \Cref{app:benchmark} \\
    $u$, $w$ & Auxiliary quantities with $10^{y}=u\,w$ (analogue of a constitutive identity) & \Cref{sec:benchmark} \\
    $\mathcal{N}(\mu,s^2)$ & Normal distribution with mean $\mu$ and standard deviation $s$ & \Cref{app:benchmark} \\
    MAE & Mean absolute error, here in decades & \Cref{sec:impact} \\
    $R^2$ & Coefficient of determination of a model's predictions & \Cref{sec:impact} \\
    \bottomrule
  \end{tabular}
\end{table}

\begin{table}[!htbp]
  \caption{Physical quantities and field-specific terms. Tensor indices
  $i,j,k,l,m$ run over the three crystal axes, and a repeated index is summed.
  Superscripts give the condition under which a quantity is defined: $T$
  constant stress (unclamped, free crystal), $S$ constant strain (clamped), $E$
  constant electric field.}
  \label{tab:glossary}
  \centering
  \small
  \begin{tabular}{L{0.20\linewidth} L{0.56\linewidth} L{0.16\linewidth}}
    \toprule
    \DDHeadRow
    \DDHead{Symbol or term} & \DDHead{Meaning} & \DDHead{Typical unit} \\
    \midrule
    \multicolumn{3}{l}{\textit{Electro-optic, dielectric, piezoelectric, and elastic tensors}} \\
    $r_{ijk}$; $r^{T}$, $r^{S}$ & Linear electro-optic (Pockels) coefficient, $\Delta(1/n^2)_{ij}=r_{ijk}E_k$; unclamped and clamped values & pm/V \\
    $r_{ij}$, $\gamma_{ij}$ & The same coefficient in Voigt notation; $\gamma$ is an alternative symbol & pm/V \\
    $r_c$ & Combined or effective coefficient that mixes several components & pm/V \\
    $E_k$ & Component $k$ of the applied electric field & V/m \\
    $n$ & Refractive index (in \cref{tab:fields,tab:glossary} and \cref{sec:consensus}) & -- \\
    $\varepsilon_{ij}$; $\varepsilon^{T}$, $\varepsilon^{S}$ & Relative permittivity (dielectric constant); unclamped and clamped values & -- \\
    $\varepsilon_0$ & Permittivity of vacuum, \SI{8.854e-12}{F/m} & F/m \\
    $d_{ijk}$, $d_{ij}$ & Piezoelectric strain coefficient: strain per unit electric field & pm/V $=$ \SI{e-12}{C/N} \\
    $e_{ijk}$, $e_{ij}$ & Piezoelectric stress coefficient: stress per unit electric field & C/m$^2$ \\
    $p^{E}_{ijlm}$ & Elasto-optic tensor: change of $1/n^2$ per unit strain & -- \\
    $c^{E}_{ijlm}$, $c_{ij}$ & Elastic stiffness at constant field (full and Voigt notation) & GPa \\
    $s^{E}_{ijlm}$ & Elastic compliance, the inverse of the stiffness & GPa$^{-1}$ \\
    Voigt notation & Compression of a symmetric index pair to one index: $11\to1$, $22\to2$, $33\to3$, $23\to4$, $13\to5$, $12\to6$ & -- \\
    $\bar{4}3m$, $4mm$, $mm2$, $3m$ & Crystal point groups in Hermann--Mauguin notation (cubic zinc blende; tetragonal; orthorhombic; trigonal) & -- \\
    $\nu$ & Poisson ratio & -- \\
    \multicolumn{3}{l}{\textit{Energy storage}} \\
    Specific capacity & Charge stored per unit mass (or area) of electrode & mAh\,g$^{-1}$ \\
    Coulombic efficiency & Charge delivered on discharge divided by charge stored on charge & \% \\
    C-rate & Charge or discharge current relative to capacity; 1C empties the cell in one hour & h$^{-1}$ \\
    Ionic conductivity & Ease of ion transport through an electrolyte & S\,cm$^{-1}$ \\
    \multicolumn{3}{l}{\textit{Medicine and drug delivery}} \\
    IC$_{50}$, pIC$_{50}$ & Concentration giving half-maximal inhibition; pIC$_{50}=-\log_{10}$IC$_{50}$ (molar) & nM, \textmu M \\
    $K_\mathrm{i}$ & Inhibition constant: dissociation constant of an inhibitor from its target & nM \\
    SD, SEM & Standard deviation; standard error of the mean, $\mathrm{SEM}=\mathrm{SD}/\sqrt{n}$ & -- \\
    Dose basis & Dose per body mass (mg\,kg$^{-1}$) or per body surface area (mg\,m$^{-2}$) & -- \\
    \multicolumn{3}{l}{\textit{Mechanics and other fields}} \\
    Engineering and true stress & Force per original area, and per current area & MPa \\
    HV & Vickers hardness number & HV, GPa \\
    ksi & Thousand pounds-force per square inch (\SI{6.895}{MPa}) & -- \\
    $T_c$ & Superconducting critical temperature & K \\
    \multicolumn{3}{l}{\textit{Governing equations of \cref{tab:equations}}} \\
    $Q_{\mathrm{th}}$, $n$, $F$, $M$ & Theoretical specific capacity; electrons transferred per formula unit; Faraday constant (\SI{96485}{C/mol}); molar mass & mAh\,g$^{-1}$ \\
    $E$, $E^{0}$, $Q_r$, $z$ & Electrode potential, its standard value, reaction quotient, and charge number (Nernst equation) & V \\
    $L$, $\Delta v$ & Latent heat and volume change of a phase transition (Clapeyron equation) & J/mol, m$^3$/mol \\
    $x_i$, $\mu_i$ & Mole fraction and chemical potential of component $i$ (Gibbs--Duhem) & --, J/mol \\
    $L_{ij}$ & Coupled transport coefficients (Onsager) & -- \\
    $\kappa(\omega)$, $\mathcal{P}$ & Extinction coefficient at angular frequency $\omega$; principal value of the integral (Kramers--Kronig) & -- \\
    $K$, $G$; $V$, $R$ & Bulk and shear modulus; subscripts for the Voigt and Reuss bounds & GPa \\
    $\sigma_y$, $D$, $k$ & Yield strength, grain size, and Hall--Petch coefficient & MPa, \textmu m \\
    $a$, $x$ & Lattice parameter and alloy composition (Vegard's law) & \AA, -- \\
    $E_a$, $k_B$, $A$ & Activation energy, Boltzmann constant, and prefactor (Arrhenius) & eV \\
    $\chi^{(1)}$, $\chi^{(2)}$ & Linear and second-order susceptibilities (Miller's rule) & --, pm/V \\
    \bottomrule
  \end{tabular}
\end{table}

\clearpage
\section{A protocol for handling erroneous data}
\label{app:protocol}

The layered approach of \cref{sec:detect} is written below as a procedure, so
that a new project can adopt it directly and the paper and the implementation
describe the same order of operations.

\begin{DDProcedure}{Erroneous-data protocol for literature-mined databases}{alg:protocol}
\PInput{Extracted records, each with a value, label, units, material,
conditions, and provenance (paper, page, table cell or sentence, raw snippet); a
controlled vocabulary; a rule library of bounds, symmetry rules, and identities
for each property family.}
\PStep{Validate the schema (L0)} Check that the label is in the vocabulary, the
units have the right dimension, the required context is present, and the value
appears in its snippet. Map notational variants to canonical symbols. Mark cited
values and remove duplicate documents. Hold failures with a typed reason.
\PStep{Apply physics (L1)} Reject values that violate hard bounds or symmetry
selection rules. For each paper or matched material group, evaluate the
available identities (\cref{eq:zphys}). When $|z_{\mathrm{phys}}|>3$, use the
pattern of violated identities to locate the suspect value and hold it.
\PStep{Build consensus (L2)} Group records by material, component, boundary
condition, and matched conditions. Compute robust scores (\cref{eq:mad}) with a
floor for scatter between laboratories, and the group consistency $R_B$. Test
flagged values for decade and convention shifts and record verified repairs as
repairs. Store the number of independent sources with each group.
\PStep{Audit with a model (L3)} Train a robust surrogate. Score every record by
its out-of-fold residual (\cref{eq:suspicion}) and, where available, by ensemble
disagreement or influence.
\PStep{Review in order of suspicion (L4)} Combine the scores into $S_i$ and
review records in decreasing order of $S_i$ within the available budget, using
the stored provenance. Feed decisions back into the rule library.
\PStep{Decide and train} Give every record one outcome: accept (with a weight),
repair, hold, down-weight, or reject. Train with confidence weights and a robust
loss. Evaluate on a cleaned held-out set.
\POutput{A database in which every value carries a status, a weight, a
corroboration count, and a provenance trail; a list of held values with reasons;
and a ranked list of acquisitions that would resolve them.}
\PNote{Nothing is deleted. A database that keeps only its accepted values cannot
be audited, and cannot tell the next acquisition step what is missing.}
\end{DDProcedure}
\normalcolor

\section{Benchmark details}
\label{app:benchmark}

\paragraph{Corpus}
Each corpus contains $M=400$ materials with descriptors $\vect{x}\in[0,1]^6$
drawn uniformly. The true property is
\begin{equation}
  y^{*} = 1.6 + 0.7\sin 3x_1 + 0.8x_2^2 - 0.6x_3 + 0.5x_4x_5 + 0.3\cos 4x_6 + \delta_m
\end{equation}
in $\log_{10}$ units, where $x_1,\dots,x_6$ are the components of $\vect{x}$ and
$\delta_m\sim\mathcal{N}(0,0.12^2)$ is a material-specific term, drawn from a
normal distribution with mean 0 and standard deviation 0.12, that the descriptors do not explain. The
number of reports of material $m$ is $n_m=1+\lfloor1.2P\rfloor$, capped at sixty,
where $\lfloor\cdot\rfloor$ rounds down and $P$ is drawn from a Lomax (Pareto
type II) distribution with shape 1.3, a heavy-tailed distribution in which most
draws are small and a few are very large. This
gives a mean of about 3.5 reports, with 54\% of materials reported once. Each
report carries a laboratory offset $\mathcal{N}(0,0.025^2)$ and a measurement
error $\mathcal{N}(0,0.035^2)$. With probability 0.4 a report also gives the
auxiliary quantities $u$ and $w$, with $\log_{10}u$ a smooth function of
$\vect{x}$ and $\log_{10}w = y^{*}-\log_{10}u$, each with its own measurement error.

\paragraph{Error injection}
Records are corrupted until a fraction $\rho$ of target values are wrong. The
error classes and their relative frequencies are: unit or scale shifts of $\pm1$
or $\pm3$ decades (0.18); OCR corruption (0.14), in which a digit of the
three-significant-figure value is replaced by a visually confusable digit or the
decimal point is dropped; label swap, in which the auxiliary $u$ is stored as the
target (0.14); misattribution to another material, half the time one of its five
nearest neighbours in descriptor space (0.14); context loss, a shift of $+0.4$ to
$+1.3$ decades mimicking a value measured near a transition (0.12);
hallucination, a value drawn from the distribution of true values across
materials with 0.1-decade jitter (0.12); and citation copying, in which one wrong
value is copied into two to four reports of a material with at least four reports
(0.16). Corruptions smaller than 0.02 decades are discarded, because they cannot
be told apart from measurement error.

\paragraph{Detectors}
Bounds: $y\notin[-0.5,4.5]$. Identity: $|y-\log_{10}u-\log_{10}w|>0.25$ decades.
Consensus: $|z_i|>3.5$ in groups with at least three reports, with
$\delta_0=0.03$. Decade repair: flagged values within 0.12 of a non-zero integer
shift from the group median. Model audit: five-fold out-of-fold predictions from
a gradient-boosted regressor with Huber loss, a loss that is quadratic for small
residuals and linear for large ones \citep{huber1964} (250 trees, depth 3, learning rate
0.06, subsample 0.8), flagging $s_i>3.5$. The layered detector is the union of
all four, and the suspicion score is the maximum of the threshold-normalised
scores.

\paragraph{Models}
In each corpus, 20\% of materials are held out. Surrogates are gradient-boosted
regressors (300 trees, depth 3, learning rate 0.05, subsample 0.8), trained on
the records of the remaining materials under six strategies: error-free records
only (reference); all records with squared loss; all records with Huber loss;
one median per material; records that survive layered cleaning, with verified
repairs applied; and all records with repairs applied and weight
$1/\max(1,S_i)^2$. MAE is computed against the true $y^{*}$ of the held-out
materials. All results are means over ten random seeds for each error rate.

\section{Details of the real-data audits}
\label{app:audits}

\paragraph{Gate reasons (\cref{tab:reasons})}
Every held or rejected record carries a free-text list of reasons written by the
verification gate. Reasons were mapped to the nine classes of
\cref{tab:reasons} by keyword, and a record is counted once in each class it
touches. One held record (a non-canonical unit string) matched no class.

\paragraph{Reference audit (\cref{tab:audit})}
All accepted records whose source is one of the three benchmark papers were
compared with the curated reference values for the same paper. Tensor labels were
converted to full-index form. Voigt and full-tensor notations, index-order
variants, and the symmetry equalities $r_{23}=r_{13}$ and $r_{42}=r_{51}$ (for
$4mm$ and $3m$) were treated as equivalent. A value matches if it lies within 2\%
(or 0.006 in absolute terms for small dimensionless values) of a reference value
with the same component and a compatible clamped or unclamped label. Decade
shifts were tested for factors of $10^{\pm1}$ to $10^{\pm3}$ with a 5\%
tolerance. For the reference-free checks, allowed tensor components were computed
by averaging basis tensors over each point group ($4mm$, $mm2$, $3m$, $6mm$,
$\bar{4}3m$), with the group chosen from material and reported phase, and records
in films or low-symmetry phases were skipped. A label was malformed if an index
was out of range for its notation. The text and numeric fields of a record had to
agree within 2\%. Diagonal stiffness constants had to lie between 20 and
800\,GPa. A within-source conflict was flagged when records sharing paper,
family, component, and boundary differed by more than 10\%.

\paragraph{Indistinguishable fraction}
$\Phi$ (\cref{eq:phi}) was computed on the 232 accepted electro-optic records of
the latest production run with $\eta=0.1$, grouping by material, canonical
component, and source paper together with the listed condition fields; records
with a missing field share the empty value for that field.

\end{document}

%% file: Preamble/DDLabIdentity.tex
\labname{Dibakar Datta Laboratory}
\labshortname{DDLab}
\institutionshort{NJIT}

\labwebsite[https://dibakardatta.net/]{www.dibakardatta.net}

\labcodeurl{https://dibakardatta.net/code/}

\citationauthors{D. Datta \emph{et al.}}

%% file: Figures/fig_pipeline.tex
\begin{figure}[tb]
  \centering
  \begin{tikzpicture}[x=1mm, y=1mm,
      every node/.append style={execute at begin node={\hyphenpenalty=10000\relax}},
      st/.style={ddmethod, text width=26mm, minimum height=19mm, inner sep=3pt},
      outp/.style={ddresult, text width=26mm, minimum height=12mm, inner sep=3pt, font=\scriptsize},
      hd/.style={font=\footnotesize\bfseries, text=DDNavy}]
    \node[st] (s0) at (14,40)  {\textbf{Stage 0}\\Acquisition\\[0.6mm]{\scriptsize search and rank papers from bibliographic services}};
    \node[st] (s1) at (47,40)  {\textbf{Stage 1}\\Extraction\\[0.6mm]{\scriptsize layout-aware parsing; LLM (GPT-4o) reads values and conditions}};
    \node[st, ddcontrol, text width=26mm, minimum height=19mm] (s2) at (80,40)
      {\textbf{Stage 2}\\Verification gate\\[0.6mm]{\scriptsize symmetry, unit, label, and provenance rules}};
    \node[st] (s3) at (113,40) {\textbf{Stage 3}\\Consensus\\[0.6mm]{\scriptsize median and MAD across papers (\cref{eq:mad})}};
    \node[st] (s4) at (146,40) {\textbf{Stage 4}\\Modelling\\[0.6mm]{\scriptsize noise-robust surrogate, leave-one-material-out audit}};
    \foreach \a/\b in {s0/s1, s1/s2, s2/s3, s3/s4} \draw[ddflow] (\a) -- (\b);

    \node[outp] (o0) at (14,17)  {49 PDFs\\(3 out of scope)};
    \node[outp] (o1) at (47,17)  {816 candidate values from 43 papers};
    \node[outp, text width=26mm] (o2) at (80,17)
      {{\color{DDGreen!70!black}\textbf{414} accepted}\\{\color{DDGold!60!black}\textbf{215} held}\enspace{\color{DDRed!80!black}\textbf{187} rejected}};
    \node[outp] (o3) at (113,17) {448 consensus records, 22 materials};
    \node[outp] (o4) at (146,17) {predicts 9 of 13 held-out materials};
    \foreach \i in {0,...,4} \draw[ddflow, draw=DDTeal!70] (s\i.south) -- (o\i.north);

    \node[ddevidence, text width=58mm, minimum height=9mm, font=\scriptsize] (gold) at (63.5,1)
      {Certified gold standard: 303 values from 16 papers, read by hand\\
       $\Rightarrow$ precision 0.51, recall 0.48 of the production run};
    \draw[ddreturn, dashed] (gold.north -| o1.south) -- (o1.south);
    \draw[ddreturn, dashed] (gold.north -| o2.south) -- (o2.south);

    \node[hd, anchor=west] at (0,53.5) {Each stage passes on only what it can support};
    \node[font=\scriptsize\itshape, text=DDInk, anchor=west] at (95,1)
      {held and rejected values are logged with a reason, never deleted};
  \end{tikzpicture}
  \caption{The electro-optic data pipeline used as our case study
  \citep{datta2026framework, datta2026scikg}. Top row: the five stages. Middle
  row: what each stage produced. The counts for Stages~0--2 come from the latest
  production run, which is analysed in \cref{sec:gate,sec:audit}; the consensus
  database and the leave-one-material-out result in Stages~3--4 summarise the
  whole campaign. The certified gold standard (bottom) is read by hand,
  independently of the pipeline, and measures how good the extracted values
  are.}
  \label{fig:pipeline}
\end{figure}

%% file: Figures/fig_fields.tex
\begin{figure}[p]
  \centering
  \begin{tikzpicture}[x=1mm, y=1mm, font=\scriptsize,
      every node/.append style={execute at begin node={\hyphenpenalty=10000\relax}},
      frame/.style={draw=DDRule, rounded corners=1.5mm, line width=0.5pt, fill=white},
      ttl/.style={anchor=north west, font=\footnotesize\bfseries, text=DDNavy},
      chk/.style={anchor=south west, text width=49mm, align=left, text=DDInk, font=\scriptsize},
      lab/.style={text=DDInk, align=center, font=\scriptsize}]

    \begin{scope}[shift={(0,62)}]
      \draw[frame] (0,0) rectangle (53,58);
      \node[ttl] at (1.5,57) {(a) Electro-optics};
      \fill[DDTeal!12] (12,26) rectangle (40,38);
      \draw[DDTeal] (12,26) rectangle (40,38);
      \fill[DDGold!65] (12,38) rectangle (40,40.2);
      \fill[DDGold!65] (12,23.8) rectangle (40,26);
      \node[lab, text=DDGold!40!black] at (26,43) {$+V$};
      \foreach \x in {16,22,...,36} \draw[-{Stealth[length=1.1mm]}, DDOrange!85] (\x,37.4) -- (\x,33.2);
      \node[lab, text=DDTeal!70!black] at (26,29) {crystal, $r_{ij}$};
      \draw[DDRed!80, line width=0.7pt, domain=2:11, samples=60, smooth] plot (\x,{32+2.2*sin(\x*120)});
      \draw[DDPurple, line width=0.7pt, domain=41:51, samples=60, smooth] plot (\x,{32+2.2*sin(\x*120-80)});
      \node[lab] at (6.5,20) {light in};
      \node[lab] at (46,20) {phase\\shifted};
      \node[chk] at (1.5,1.5) {\textbf{Mined:} Pockels coefficient $r_{ij}$ (pm/V), clamped or unclamped.\\
        \textbf{Check:} symmetry allows only some $r_{ij}$; clamped and unclamped values obey \cref{eq:identities}.};
    \end{scope}

    \begin{scope}[shift={(56,62)}]
      \draw[frame] (0,0) rectangle (53,58);
      \node[ttl] at (1.5,57) {(b) Energy storage};
      \fill[DDGray!35] (8,20) rectangle (18,42);  \draw[DDGray] (8,20) rectangle (18,42);
      \fill[DDNavy!15] (35,20) rectangle (45,42); \draw[DDNavy!60] (35,20) rectangle (45,42);
      \draw[DDMeta, dashed] (26.5,20) -- (26.5,42);
      \node[lab] at (13,17) {anode};
      \node[lab] at (40,17) {cathode};
      \node[lab] at (26.5,17) {separator};
      \foreach \y in {26,31,36} {
        \node[circle, fill=DDTeal!70, inner sep=0pt, minimum size=1.8mm] at (22,\y) {};
        \draw[-{Stealth[length=1.1mm]}, DDTeal] (23.5,\y) -- (31,\y);
      }
      \node[lab, text=DDTeal!70!black] at (26.5,40) {Li$^{+}$};
      \draw[DDInk, line width=0.5pt] (13,42) -- (13,47) -- (40,47) -- (40,42);
      \draw[-{Stealth[length=1.3mm]}, DDOrange!85, line width=0.6pt] (22,47) -- (31,47);
      \node[lab, anchor=south] at (26.5,47.4) {e$^{-}$ through the load};
      \node[chk] at (1.5,1.5) {\textbf{Mined:} specific capacity (mAh\,g$^{-1}$), cycle life, conductivity.\\
        \textbf{Check:} capacity cannot exceed the Faraday limit $nF/(3.6M)$ of the active material.};
    \end{scope}

    \begin{scope}[shift={(112,62)}]
      \draw[frame] (0,0) rectangle (53,58);
      \node[ttl] at (1.5,57) {(c) Drug delivery};
      \fill[DDPurple!12] (15,32) circle (7mm); \draw[DDPurple!70] (15,32) circle (7mm);
      \foreach \a in {0,45,...,315} {
        \draw[DDPurple!70] ($(15,32)+(\a:7mm)$) -- ($(15,32)+(\a:9.5mm)$);
        \fill[DDPurple!70] ($(15,32)+(\a:9.9mm)$) circle (0.5mm);
      }
      \foreach \p in {(13,34),(17,35),(15,30),(12,30),(18,31),(15,33.5)}
        \fill[DDRed!75] \p circle (0.7mm);
      \node[lab] at (15,18) {nanoparticle\\carrying drug};
      \draw[DDGreen!60!black, line width=0.9pt] (38,16) .. controls (34,26) and (34,38) .. (38,48);
      \foreach \y in {24,32,40} \draw[DDGreen!60!black, line width=0.6pt] (35.2,\y) -- (32.5,\y) -- (31.8,\y+1) (32.5,\y) -- (31.8,\y-1);
      \node[lab, text=DDGreen!50!black] at (45,32) {target\\cell};
      \draw[-{Stealth[length=1.3mm]}, DDInk] (25.6,32) -- (30.6,32);
      \node[chk] at (1.5,1.5) {\textbf{Mined:} encapsulation efficiency, release, IC$_{50}$ (nM or \textmu M).\\
        \textbf{Check:} efficiencies lie in 0--100\%; $\mathrm{SEM}=\mathrm{SD}/\sqrt{n}$.};
    \end{scope}

    \begin{scope}[shift={(0,0)}]
      \draw[frame] (0,0) rectangle (53,58);
      \node[ttl] at (1.5,57) {(d) Mechanics};
      \draw[-{Stealth[length=1.3mm]}, DDInk] (8,18) -- (49,18) node[below left, lab] {strain};
      \draw[-{Stealth[length=1.3mm]}, DDInk] (8,18) -- (8,50) node[below right, lab] {stress};
      \draw[DDNavy, line width=0.8pt] (8,18) -- (14,38) .. controls (18,42) and (30,47) .. (36,46) .. controls (40,45) and (43,41) .. (45,36);
      \fill[DDOrange] (14,38) circle (0.8mm);
      \node[lab, anchor=west, text=DDOrange!60!black] at (15,36) {yield};
      \fill[DDRed!75] (36,46) circle (0.8mm);
      \node[lab, anchor=south, text=DDRed!70!black] at (36,46.8) {ultimate};
      \draw[DDTeal!80, dashed, line width=0.6pt] (36,46) .. controls (40,48) and (43,49) .. (46,49.5);
      \node[lab, anchor=west, text=DDTeal!70!black] at (46.3,49.5) {true};
      \node[chk] at (1.5,1.5) {\textbf{Mined:} yield and ultimate strength, hardness.\\
        \textbf{Check:} ultimate $\ge$ yield; engineering and true stress must not be mixed.};
    \end{scope}

    \begin{scope}[shift={(56,0)}]
      \draw[frame] (0,0) rectangle (53,58);
      \node[ttl] at (1.5,57) {(e) Superconductors};
      \draw[-{Stealth[length=1.3mm]}, DDInk] (8,18) -- (49,18) node[below left, lab] {temperature $T$};
      \draw[-{Stealth[length=1.3mm]}, DDInk] (8,18) -- (8,50) node[below right, lab] {resistance};
      \draw[DDNavy, line width=0.8pt] (8,18.2) -- (22,18.2) .. controls (26,18.2) and (27,42) .. (31,43) -- (47,46);
      \foreach \x/\c/\t in {23/DDTeal/zero, 26.2/DDGold/mid, 30/DDRed/onset} {
        \draw[\c, dashed] (\x,18) -- (\x,48);
      }
      \node[lab, anchor=south, text=DDTeal!70!black] at (20.5,48) {zero};
      \node[lab, anchor=south, text=DDGold!50!black] at (26.2,48) {mid};
      \node[lab, anchor=south, text=DDRed!70!black] at (32.5,48) {onset};
      \node[chk] at (1.5,1.5) {\textbf{Mined:} critical temperature $T_c$, critical field.\\
        \textbf{Check:} the three criteria give different $T_c$; the criterion and pressure must be recorded.};
    \end{scope}

    \begin{scope}[shift={(112,0)}]
      \draw[frame] (0,0) rectangle (53,58);
      \node[ttl] at (1.5,57) {(f) Thermophysics};
      \draw[-{Stealth[length=1.3mm]}, DDInk] (8,18) -- (49,18) node[below left, lab] {temperature $T$};
      \draw[-{Stealth[length=1.3mm]}, DDInk] (8,18) -- (8,50) node[below right, lab] {pressure $P$};
      \draw[DDNavy, line width=0.8pt] (12,20) .. controls (25,23) and (35,32) .. (44,45);
      \fill[DDNavy] (44,45) circle (0.8mm);
      \node[lab, anchor=west] at (38,48) {critical point};
      \node[lab, text=DDTeal!70!black] at (18,40) {liquid};
      \node[lab, text=DDOrange!60!black] at (42,23) {vapour};
      \draw[DDRed!75, line width=0.6pt] (26.6,25.3) -- (38.6,36.6);
      \fill[DDRed!75] (32.6,31.45) circle (0.6mm);
      \node[lab, text=DDRed!70!black, anchor=north west] at (36.5,32.5) {slope $\mathrm{d}P/\mathrm{d}T$};
      \node[chk] at (1.5,1.5) {\textbf{Mined:} vapour pressure, density, latent heat.\\
        \textbf{Check:} Clapeyron, $\mathrm{d}P/\mathrm{d}T = L/(T\Delta v)$, ties the slope to the latent heat.};
    \end{scope}
  \end{tikzpicture}
  \caption{Six fields in which AI models rely on values mined from papers
  (\cref{tab:fields}). Each panel shows the physics behind the mined quantity
  and one check that exposes a wrong value without any comparison data.
  (a)~In an electro-optic crystal an applied voltage changes the refractive
  index and so shifts the phase of light passing through it (Pockels effect;
  \citealp{nye1985, yariv1984optical}).
  (b)~In a lithium-ion cell, lithium ions move through the electrolyte while
  electrons flow through the external circuit; the charge the active material
  can store sets an upper limit on the specific capacity (Faraday's law;
  \citealp{goodenough2013}).
  (c)~A drug-carrying nanoparticle with surface ligands binds to receptors on a
  target cell \citep{mitchell2021}; encapsulation and release fractions are
  bounded by 0 and 100\%.
  (d)~A tensile test gives yield and ultimate strengths; engineering stress
  (force over original area) and true stress (force over current area) differ
  at large strain \citep{borg2020}.
  (e)~The superconducting transition has a width, so the onset, midpoint, and
  zero-resistance criteria give different values of $T_c$ \citep{stanev2018}.
  (f)~On a phase diagram the Clapeyron equation links the slope of the
  liquid--vapour line to the latent heat $L$ and the volume change $\Delta v$
  \citep{callen1985}.}
  \label{fig:fields}
\end{figure}

%% file: Figures/fig_consensus.tex
\begin{figure}[tb]
  \centering
  \begin{tikzpicture}[x=1mm, y=1mm, font=\scriptsize,
      every node/.append style={execute at begin node={\hyphenpenalty=10000\relax}},
      ok/.style={circle, fill=DDGreen!70, draw=DDGreen!50!black, inner sep=0pt, minimum size=2.6mm},
      bad/.style={circle, fill=DDRed!75, draw=DDRed!50!black, inner sep=0pt, minimum size=2.6mm},
      dec/.style={circle, fill=DDOrange!80, draw=DDOrange!50!black, inner sep=0pt, minimum size=2.6mm},
      lab/.style={align=center, text=DDInk}]
    \fill[DDGreen!10] (21.5,4) rectangle (46.5,30);
    \node[lab, text=DDGreen!50!black, anchor=east] at (20.5,26) {accepted\\band,\\$|z_i|\le3.5$};
    \draw[DDInk, line width=0.6pt, -{Stealth[length=2mm]}] (4,4) -- (146,4);
    \foreach \x/\t in {26/1.5, 66/2.0, 106/2.5} {
      \draw[DDInk] (\x,4) -- (\x,2.8);
      \node[anchor=north, text=DDInk] at (\x,2.6) {\t};
    }
    \node[anchor=north, text=DDInk] at (134,2.6) {$y=\log_{10}|v|$};

    \draw[DDNavy, line width=0.9pt] (34.8,4) -- (34.8,26);
    \node[lab, text=DDNavy, anchor=east] at (34.3,24) {median $\tilde y$};
    \draw[DDGray!80!black, line width=0.8pt, dashed] (47.6,4) -- (47.6,24);
    \node[lab, text=DDGray!30!black, anchor=west] at (48.3,24) {mean, pulled by the two errors};

    \foreach \x/\h in {30/8, 32.4/12, 34/8, 34.8/16, 35.6/12, 37.2/8, 38.8/12}
      \node[ok] at (\x,\h) {};
    \node[bad] at (70,8) {};
    \node[lab, anchor=south, text=DDRed!70!black] at (70,10.5)
      {random error,\\flagged ($|z_i|\gg3.5$)};
    \node[dec] at (114.8,8) {};
    \node[lab, anchor=south, text=DDOrange!60!black] at (114.8,10.5)
      {unit slip, $\Delta y\approx1$\\(a factor of ten)};
    \draw[-{Stealth[length=2mm]}, DDOrange!85, line width=0.7pt, dashed]
      (114.8,19) .. controls (104,46) and (56,46) .. (40,29);
    \node[lab, text=DDOrange!60!black, anchor=south] at (80,40.5)
      {proposed repair: $y_i-1$, stored as a hypothesis};
  \end{tikzpicture}
  \caption{Robust consensus in one group of reports of the same quantity
  (schematic). Correct values (green) cluster around the true value, spread only
  by the differences between laboratories. The median (solid line) stays inside
  the cluster, whereas the mean (dashed) is pulled towards the two errors. The
  robust score of \cref{eq:mad} measures each report's distance from the median
  in units of the median absolute deviation. A report far outside the band is
  flagged (red); one that sits almost exactly a whole number of decades away
  (orange) is most likely a unit or decimal slip, and the shifted value is
  stored as a proposed repair, never written over the raw value.}
  \label{fig:consensus}
\end{figure}